\documentclass[showpacs,aps,pre,twocolumn]{revtex4-1}
\usepackage[utf8]{inputenc}
\usepackage{amsmath}
\usepackage{amssymb}
\usepackage{siunitx}
\usepackage{microtype}
\usepackage[english]{babel}
\usepackage{bm}
\usepackage{color}
\usepackage[capitalise]{cleveref}
\usepackage{etoolbox}
\usepackage{appendix}

\makeatletter
\appto{\appendix}{%
	\@ifstar{\def\theequation@prefix{A.}}%
	{}%
}
\makeatother

\newcommand{\eps}{\frac{1}{\epsilon}}

\newcommand{\be}{\mathbf{e}}
\newcommand{\bx}{\mathbf{x}}
\newcommand{\bp}{\mathbf{p}}
\newcommand{\bs}{\mathbf{s}}
\newcommand{\bE}{\mathbf{E}}
\newcommand{\bB}{\mathbf{B}}
\newcommand{\bsigma}{\bm{\sigma}}

\newcommand{\bj}{\mathbf j}

\DeclareMathOperator{\Tr}{Tr}

\begin{document}
	
\title{Relativistic kinetic theory for spin-1/2 particles: Conservation laws, thermodynamics, and linear waves} 
\author{R. Ekman, H. Al-Naseri, J. Zamanian, and G. Brodin}
\affiliation{Department of Physics, Ume{\aa } University, SE--901 87 Ume{\aa}, Sweden}

\begin{abstract}
    We study a recently derived fully relativistic kinetic model for spin-$1/2$ particles.
	Firstly, the full set of conservation laws for energy, momentum and angular momentum are given, together with an expression for the (non-symmetric) stress-energy tensor.
	Next, the thermodynamic equilibrium distribution is given in different limiting cases.
	Furthermore, we address the analytical complexity that arises when  the spin- and momentum eigenfunctions are coupled in linear theory, by calculating the linear dispersion relation for such a case.
	Finally, we discuss the model and give some context by comparing with potentially relevant phenomena that are not included, such as radiation reaction and vacuum polarization.   
\end{abstract} 

\maketitle

\section{Introduction}

Current and next-generation high-intensity laser facilities provide opportunities to study highly relativistic electron dynamics.
Here the electron spin~\cite{Hu1999,Walser2000,Walser2002,Wen2016} is of interest, as the electrons may spin-polarize~\cite{PhysRevA.96.043407,delSorbo2018polarization,PhysRevLett.122.154801}, in just a few laser cycles.
The ultra-strong magnetic fields present in astrophysical environments~\cite{Harding2006,uzdensky2014extreme,Mahajan2014} also enhance the significance of spin-related phenomena.

In less extreme settings, applications in, for example, spintronics~\cite{wolf2001spintronics},  quantum wells~\cite{Manfredi-quantum-well}, and plasmonics~\cite{atwater2007plasmonics} have also stimulated an interest in high density plasmas where quantum effects are significant~\cite{Shukla-Eliasson-RMP,Haas-book}.

Earlier spin-kinetic models have included effects such as the magnetic dipole force, magnetization currents, and spin precession~\cite{PhysRevLett.101.245002,ZamanianNJP,asenjo2012semi,hurst2014semiclassical,andreev2015quantum,andreev2017kinetic}, but have been limited to particle velocities well below the speed of light in vacuum. 
However, in a previous paper~\cite{PhysRevE.96.023207}, which we will refer to as Paper~I below, a fully relativistic kinetic equation for spin-$\frac{1}{2}$ was presented, along with its couplings to Maxwell's equations.
This forms a model that can be used to describe self-consistent relativistic plasma dynamics, including spin effects.

Here, our purpose is to study some basic properties of the model, and also provide some tools needed for analytical studies. The organization  of the paper is as follows: 
In~\cref{sec:overview} we give an overview of the model, including the assumptions made in the derivation.
Next, in~\cref{sec:conservation}, we derive conservation laws for energy, momentum, and angular momentum and give an expression for the stress energy tensor. 
It turns out that the stress-energy tensor is non-symmetric in our case, which is related to the presence of  spin angular momentum. 
\cref{sec:thermodynamics} is devoted to a study of the thermodynamic background distribution for various cases and the associated background magnetization due to the electrons. 
In particular we present the expression for the relativistic and the non-relativistic case allowing for, in both cases, non-degenerate and degenerate electrons. 
Next, in~\cref{sec:disprel}, we analyze linear theory in a homogeneous magnetized medium. 
It turns out due to relativistic effects the spin- and momentum eigenfunctions become coupled, and the standard solution procedure must be generalized. 
We give an example of how this can be done, and present a dispersion relation in the limiting case of wave propagation perpendicular to the magnetic field.  
Finally, in~\cref{sec:discussion}, the significance of our results and the applicability of the model are discussed.

\section{Overview of the model}
\label{sec:overview}

The model comes from separating positive and negative energy solutions of the Dirac equation by means of a Foldy-Wouthuysen (F-W) transformation~\cite{foldy1950dirac,Silenko2008}.
Since we are decoupling electrons and positrons, the physical condition of applicability is that pair production is negligible.
Quantitatively, the fields should not be comparable to the critical fields, $E \ll E_c = m^2/q \hbar$ and similarly for $B$, and their typical scale lengths should be long compared to the Compton wavelength $\hbar/m$.
We then take a gauge-invariant Wigner transformation~\cite{wigner1932,stratonovich1956gauge,serimaa} (see Ref.~\cite{case2008wigner} for a pedagogical introduction to the Wigner formalism) to obtain an evolution equation for a $2\times 2$-matrix valued Wigner function $W_{\alpha\beta}(\mathbf{x}_i, \mathbf{p}_i, t)$.
Here $\mathbf{x}_i, \mathbf{p}_i$ denote that in a many-body system the Wigner function depends on all the positions and momenta.
However, the BBGKY hierarchy applies, and neglecting collisions the evolution equation for the one-particle Wigner function is found from the one-particle Hamiltonian.

Next, we apply a spin transformation~\cite{ZamanianNJP},
	\begin{align}
		f(\bx, \bp , \bs , t) & :=  
		\frac{1}{4\pi} \textrm{Tr} 
			\left[ ( 1 + \bs \cdot \bm \bsigma) W(\bx, \bp) \right] \notag \\ &
			= \frac{1}{4\pi}  \left( \delta_{\alpha \beta} + \bs \cdot \bm \bsigma_{\alpha \beta} 
		\right) W_{\beta \alpha} ( \bx, \bp , t ) , 
\end{align}
to obtain a scalar Wigner function (summation convention applied to the spin indices).
In this formalism, densities in space are given by moments of the Wigner function over the momentum and spin variables $\bp$ and $\bs$.
For example, the number density is $n = \int d^3 p d^2s \,  f $.
Below, we will often use the notation $d\Omega = d^3p d^2s$ and \begin{equation}
	\langle \Phi \rangle := \int d\Omega \, \Phi f
\end{equation}
where $\Phi$ is some function on phase space.
The normalization of the spin transformation is such that the spin density is \begin{equation}
	\langle \bx | \bsigma_{\alpha \beta} \rho_{\beta \alpha} | \bx \rangle = 
	3 \int d\Omega \, \mathbf{s} f.
\end{equation}
Also, it is easy to derive the identities
\begin{align} 
	\int d \Omega \, \nabla_s  f  & = 2 \int d \Omega \, \bs f 
	\label{eq:spinIdentityI}  \\
	\int d \Omega \, \bs \times \mathbf X \cdot \nabla_s f & = 0 
	\label{eq:spinIdentityII}
\end{align} 
where $\mathbf X$ is any vector independent of $\mathbf s$.
These identities are used repeatedly below.

The evolution equation for the scalar Wigner function $f$ is 
\begin{widetext} 
	\begin{align}
	0 &  = \partial_t f + 
	\left( 
		\frac{\bp}{\epsilon} - \mu_B \nabla_p \tilde{T}
	\right) 
	\cdot \nabla_x f 	
	+ q \left(
		\bE
		+ \left( \frac{\bp}{\epsilon} - \mu_B  \nabla_p \tilde{T} \right)
		\times \mathbf B
	\right)
	\cdot \nabla_p f
	+ \mu_B (\nabla_x \tilde{T} ) \cdot \nabla_p f
	+ \frac{2 \mu_B m }{\hbar \epsilon}  \left(
		\bs \times \mathbf T
	\right)
	\cdot \nabla_s f
	\label{eq:evolution}
	\end{align} 
\end{widetext} 
where $\epsilon^2 = \bp^2 + m^2$, $\mu_B = q \hbar / 2m$ is the Bohr magneton and \begin{align}
\mathbf{T} & = \frac{m}{\epsilon} \left( \bB - \frac{\bp \times \bE}{\epsilon + m} \right) \\
\tilde{T}  & = \mathbf T\cdot (\bs + \nabla_s).
\end{align}

The system is closed with Maxwell's equations, in units where $ c = \epsilon_0 = \mu_0 = 1 $, 
\begin{subequations}
	\begin{align}
	\nabla\cdot\mathbf E & = \rho_f - \nabla\cdot\mathbf P \\ 
	\nabla \cdot \mathbf B & = 0 \\
	\nabla \times \mathbf E & = - \frac{\partial \mathbf B}{\partial t} \\
	\nabla\times \mathbf B & = \mathbf j_f + \frac{\partial \mathbf E}{\partial t} + \frac{\partial \mathbf P}{\partial t} + \nabla\times \mathbf M \label{eq:ampere}
	\end{align}
\end{subequations}
where $\mathbf P$ and $\mathbf M$ are the polarization and magnetization densities, and $\rho_f$ and $\mathbf j_f$ are the free charge and current densities. 
These are given by 
\begin{align}
	\rho_f  & = q \int d\Omega \, f \\
	\mathbf j_f & = q\int d\Omega \, \left(
		\frac{\mathbf p}{\epsilon}
		- \mu_B \nabla_p \mathbf T	\cdot 3\mathbf s
	\right)f \label{eq:j-free}\\
	\mathbf P  & = -3\mu_B  \int d\Omega \, \frac{m\mathbf s \times \mathbf p}{\epsilon(\epsilon+m)} f \label{eq:polarization} \\
	\mathbf M & = 3\mu_B \int d\Omega\, \frac{m}{\epsilon} \mathbf s f \label{eq:magnetization}
\end{align}
and we refer to Paper~I for the derivation.
It follows from the evolution equation, \cref{eq:evolution}, that the free charge is conserved, $\partial_t \rho_f + \nabla \cdot \mathbf j_f = 0$, and we interpret \begin{equation}
	\mathbf v = \frac{\mathbf p}{\epsilon} - \mu_B \nabla_p \mathbf T	\cdot 3\mathbf s
	\label{eq:velocity}
\end{equation}
as the function on phase space corresponding to the velocity -- it is in fact the Weyl transform of the velocity operator $\hat{\mathbf v} = \frac{i}{\hbar} [\hat{H}, \hat{\bx}]$ given by the Heisenberg equation of motion.
The spin-dependent term is related to the ``hidden momentum''~\cite{PhysRevLett.18.876,PhysRevLett.20.343,PhysRev.171.1370,Babsonetal2009} of systems with magnetic moments.
It is discussed further in Paper~I, with additional references. Here we make some further observations: 

\paragraph{Anomalous magnetic moment}
The derivation of the model considered here is based on the Dirac theory where the gyromagnetic ratio, $g = 2$, discarding the anomalous magnetic moment (AMM). 
However, making this assumption is not necessary for the Foldy-Wouthuysen transformation~\cite{Silenko2008}, and we will allow for $ g \neq 2 $ below. 
Let us discuss the validity of this.

Including the AMM corresponds to adding a term $(g-2)F_{\mu\nu} \overline{\psi} \sigma^{\mu\nu} \psi $ to the Dirac equation and consequently new terms will appear in the Hamiltonian in the F-W representation, rather than just modifying $g$.
One of those new terms has the form $(g-2)\bsigma \cdot \bB$, i.e., unlike the Dirac magnetic moment, there is no $1/\varepsilon$ length contraction factor.
Since $g-2 \approx \frac{\alpha}{\pi}$ is small, the modification $g \mapsto g' = 2 + \alpha/\pi$ should be satisfactory for a phenomenological description.
Non-relativistically, such an analysis was carried out in Ref.~\cite{PhysRevLett.101.245002} and later more generally in Ref.~\cite{lundin2010linearized}, finding new wave-particle resonances due to the mismatch between the electron cyclotron and the Larmor precession frequencies.
We will continue in this spirit and consider $g -2 \neq 0$ in~\cref{sec:disprel}. 

\paragraph{Radiation reaction}
As it stands, the model does not include the effect of radiation reaction (RR, reviewed in, e.g., Refs.~\cite{DiPiazza2012,Burton2014}).
In principle, a radiation reaction force could be added to~\cref{eq:evolution} as a $\mathbf{F}_\text{RR} \cdot \nabla_p f$ term, possibly including the spin-dependence of the RR force~\cite{Bhabha1941}.
It is also now well known how to include RR in a particle-in-cell scheme~\cite{gonoskov2015extended}.
However, the ratio of the RR to the Lorentz force is $\eta = \alpha \gamma^2 E/E_\text{crit}$ where $\gamma$ is the Lorentz factor and $\alpha$ the fine-structure constant.
The strong RR regime $\eta \approx 1$ is only expected to be reached with next-generation \SI{10}{PW} laser facilities~\cite{DiPiazza2012} (but see Refs.~\cite{PhysRevX.8.011020,PhysRevX.8.031004} for a recent experiment using the Gemini laser), and by inspection
there are clearly regimes where $\eta \ll 1$, but $\gamma$ is large enough that an $O(v^2/c^2)$ treatment is inapplicable.
Furthermore, even for strong fields, the spin-dependent non-RR forces may be comparable to the RR force or even dominate it~\cite{Mahajan2014}.
However, the present model is not developed exclusively for strong fields and relativistic spin effects could be important, e.g., for relativistic temperatures (thermodynamic or Fermi).


\section{Conservation laws and stress-energy tensor}
\label{sec:conservation}

As a sanity check, the model should have conservation laws for energy, momentum, and angular momentum.
In the previous paper, the conservation law for energy was given and discussed in relation to the Abraham-Minkowski dilemma; here we give the full set of conservation laws.

\subsection{Energy and momentum}

The total energy density is given by \begin{equation}
W = \frac{1}{2}(E^2 + B^2) + \int d \Omega \, \left (\epsilon - 3\mu_B m \frac{\mathbf B}{\epsilon} \cdot \mathbf s\right)f
\end{equation}
and with the energy flux vector
\begin{multline}
	\mathbf K =
	\int d \Omega \, \left[
		\epsilon + \mu_B m 3\mathbf s \cdot \left(\frac{\mathbf B}{\epsilon} - \frac{\mathbf p\times \mathbf E}{\epsilon(\epsilon+m)}\right)
	\right] \mathbf v  f 
	\\+  \mathbf E \times \mathbf H , 
\label{eq:energy-flux}
\end{multline}
where $ \mathbf{H} = \mathbf{B} - \mathbf{M} $, 
we have a conservation law on divergence form 
\begin{equation}
	\partial_t W + \nabla\cdot \mathbf K = 0 
\label{eq:conservation-energy}.
\end{equation}

To derive the conservation of momentum, we look at 
\begin{equation}
	\partial_t \langle \bp \rangle = \partial_t \int d \Omega \, \bp f  
	= \int d \Omega \, \bp \partial_t f .
\end{equation}
We substitute for $\partial_t f$ using the kinetic equation.

Because $\bp$ is independent of $\bs$, we can divide the integral as $\int d^3p \, \bp \int d^2s \, \partial_t f$
and using the spin integral identities \cref{eq:spinIdentityI,eq:spinIdentityII} we arrive at 
\begin{multline}
- \int d \Omega \, \bp \partial_t f =
\int d^3 p \,  \bp \ \int d^2 s \, \big[ \mathbf v \cdot \nabla_x f  \\ 
+ q(\bE + \mathbf v \times \bB) \cdot \nabla_p f 
+ 3\mu_B (\bs \cdot \partial_{x_i} \mathbf T) \partial_{p_i} f  \big] . 
		\label{eq:mc-spin-integral}
\end{multline} 
The first term here is $\int d \Omega \, p_i v_j \nabla_{x_j} f$, however, the $x$-divergence \emph{cannot} be taken outside the integral to write this as the divergence of a moment, because $\mathbf v$ depends on $\mathbf x$ through $\mathbf T$.

In the second term we integrate by parts.
Since $\partial_{p_j} v_k$ is symmetric in its indices, we have 
\begin{multline}
	q\int d \Omega \, p_i (E_j + \epsilon_{jkl} v_k B_l) \partial_{p_j} f \\
	= -q\int d \Omega \, \delta_{ij}\left(E_j + \epsilon_{jkl}  v_k B_l \right) f \,\\
	= -\left(\rho_\text{f} \bE + \bj_\text{f} \times \mathbf B\right)_i
\end{multline}
which is the Lorentz force density. The spin-dependent part will be found below.

For the third term in \cref{eq:mc-spin-integral}, we have
\begin{multline}
3\mu_B \int d \Omega \, p_i (s_k \cdot \partial_{x_j} T_k) \partial_{p_j} f 
\\ 
= -3\mu_B \int d \Omega \, (\delta_{ij} s_k \partial_{x_j} T_k + p_i s_k \partial_{x_j} \partial_{p_j} T_k )f \\
=  \int d \Omega \, (-3\mu_Bs_k \partial_{x_i} T_k + p_i \partial_{x_j} v_j ) f .
\label{eq:spin-force}
\end{multline} 
For the first term above, it is simple to establish that \begin{equation}
	3\mu_B \int d \Omega \, s_k (\partial_{x_j} T_k) f = M_k \partial_{x_j} B_k 
	+ P_k \partial_{x_j} B_k
	\label{eq:gradT-scalar-s}
\end{equation}
using the definitions of $\mathbf{M}, \mathbf{P}$ in \cref{eq:magnetization,eq:polarization}.
The second term in \cref{eq:spin-force}, containing $\partial_{x_j} v_j$, is what we need for the divergence of a moment with an $\mathbf{x}$-dependence.
It is of course not a miracle that this happens, but a consequence of that the kinetic equation essentially has the form of a Poisson bracket between $f$ and the Hamiltonian.

\begin{widetext}
To summarize,
	\begin{multline}
	\partial_t \langle p_i \rangle 
	=
	-\int d \Omega \, \left [ p_i v_j \partial_{x_j}  + p_i (\partial_{x_j} v_j)  
	- 3\mu_B s_j (\partial_{x_i} T_j) \right] f  
	+ \left(\rho_\text{f} \bE + \bj_\text{f} \times \mathbf B\right)_i \\
	= 
	-\partial_{x_j} \int d \Omega \, p_i v_j  f 
	+ \left(\rho_\text{f} \bE 
	+ \bj_\text{f} \times \mathbf B\right)_i 
	+ M_j \partial_{x_i} B_j  + P_j \partial_{x_i} E_j 
	\label{eq:p-transport}
\end{multline}
using, on the second line, \cref{eq:gradT-scalar-s}.
Here we can identify 
\begin{equation}
	T_{ij}^\text{e} = \langle p_i v_j \rangle 
\end{equation}
as the stress tensor for the electrons.
Note that the relation between $\mathbf{p}$ and $\mathbf{v}$ also contains field variables, and there will be no clean separation of the total stress tensor into ``field'' and ``particle'' parts.
This is what one would expect for an interacting theory.
\end{widetext}

We now need to find an appropriate Poynting vector and electromagnetic stress tensor to match this force density.
From Maxwell's equations, using various vector identities, one finds that 
\begin{multline}
	\partial_t (\mathbf D \times \mathbf B)_i + \partial_{x_j} T_{ij}^\text{EM} \\ 
	= - \left( \rho_f \mathbf{E} + \mathbf j_f \times \bB  \right)_i - M_k \partial_{x_i} B_k 
	- P_k \partial_{x_i} E_k
\end{multline}
where $ \mathbf{D} = \mathbf{E} + \mathbf{P} $, and the tensor $T_{ij}^\text{EM}$ is defined by 
\begin{equation}
 T_{ij}^\text{EM} = \frac{1}{2} \left( E^2 + B^2 - 2 \mathbf M \cdot \bB \right) \delta_{ij} - H_i B_j - E_i D_j.
\end{equation}
where the source term is precisely the negative of that in~\cref{eq:p-transport}.
We should note that while we have used notation to indicate that this is the ``field'' part of the stress tensor, it still contains particle variables through the magnetization and polarization.

Thus, we can express conservation of momentum as 
\begin{equation}
	\partial_t \left( \langle \bp \rangle 
	+ \mathbf D \times \bB \right)_i 
	+ \partial_{x_j} (T_{ij}^\text{e} + T_{ij}^\text{EM})	 = 0.
\end{equation}

The stress tensor we have found is not symmetric, and because it is $\bE \times \mathbf H$ that appears in the energy conservation law, the full stress-energy tensor is not symmetric under $0i \leftrightarrow i0$.
This is usually considered a defect, for two reasons, discussed in more detail in Landau and Lifshitz vol.~2~\cite{landau1987classical}, §32--33 and §94.

The first reason is that to conserve angular momentum $\mathbf r \times \bm{\Pi}$, the stress tensor should be symmetric.
However, this condition applies only if the angular momentum is entirely \emph{orbital} angular momentum.
If there is spin angular momentum, angular momentum can be conserved even with a stress tensor that is not symmetric, and we shall demonstrate explicitly that this is the case for our model, below.

Identifying the correct stress-energy tensor is a general problem for field theories~\cite{forger2004currents}, since to a tensor $T_{\mu\nu}$ with $\partial_\mu T^\mu{}_\nu = 0$, we can add any 4-divergence $\partial_\rho \psi^\rho{}_{\mu\nu}$ with $\psi_{\rho\mu\nu}$ anti-symmetric in $\rho\mu$ to obtain a new tensor with vanishing 4-divergence.
Hence, the integrals $\int d^3x \, T^{i0}$ -- the actual conserved quantities -- will be unchanged, but their densities in space are not uniquely defined without a further physical principle.
As a familiar example, indeed, the tensor obtained by applying Noether's theorem to the free electromagnetic field is neither symmetric nor even gauge invariant, but can be made so~\cite{[][{, Ch.~12.10}.]Jackson}.
Still, merely requiring the tensor to be symmetric may not be enough to guarantee uniqueness~\cite{forger2004currents}.

The physical principle that ensures uniqueness is that the components of the stress-energy tensor are in principle observable: one needs simply a ruler precise enough to measure the curvature of spacetime.
This is the second reason to prefer a symmetric stress-energy tensor: it is the source of gravitation, and the left-hand side of Einstein's field equations is the symmetric Einstein tensor.
This also provides a method to obtain the stress-energy tensor directly: it is the variation of the matter action with respect to the metric. 

For manifestly Lorentz-covariant theories, the Bel\-in\-fante-Rosen\-feld construction~\cite{belinfante1940current,rosenfeld1940tenseur} gives an explicit form for $\psi_{\rho\mu\nu}$ to symmetrize the Noether tensor, but this construction relies on the structure of the Lorentz group.
Because our model is in Hamiltonian form and written in terms of $\bE, \bB$, we have broken manifest covariance and cannot use the Belinfante-Rosenfeld construction directly.

The Belinfante-Rosenfeld method, or the variation with respect to the metric, can be applied to the manifestly covariant Lagrangian in the Dirac representation, yielding a symmetric tensor $\overline{\psi} \hat{T}_{\mu\nu}^\text{D} \psi = \overline{\psi} (\gamma_\mu D_\nu +  \gamma_\nu D_\mu) \psi$, $D_\mu$ being the gauge covariant derivative.
An alternative approach would therefore be to transform the operator $\hat{T}_{\mu\nu}^\text{D}$ to the Foldy-Wouthuysen representation.
However, there are issues with gauge invariance to consider~\cite{PhysRevD.15.1063}, and this is outside the scope of the present paper.

\subsection{Spin transport}

To establish conservation of angular momentum, we consider $\partial_t \langle \bs \rangle$.
Here we must establish a few facts about the spin moments.
Because products of Pauli matrices are reducible to Pauli matrices, there are no higher spin moments.
That is, because $\sigma_i \sigma_j = \delta_{ij} + i\epsilon_{ijk} \sigma_k$, we must have $\int s_i s_j f \, d^2 s \propto \delta_{ij}$ since moments of the Wigner function should correspond to symmetrically ordered operators.
\newcommand{\normord}[1]{:\mathrel{#1}:}
In calculating the spin transport, we will want the moment corresponding to the \emph{operator} $\normord{\sigma_i \hat{v}_j}$ where the colons indicate Weyl ordering.
Using the Pauli matrices relation, we find that 
\begin{equation}
\normord{\sigma_i \hat{v}_j} = \sigma_i \frac{\hat{p}_j}{m} - \mu_B \partial_{p_j} \hat{T}_i.
\end{equation}

Because $f = \frac{1}{4\pi} \Tr [(1 + \bs \cdot \bm{\sigma}) W]$, the moment corresponding to this operator is 
\begin{equation}
	\langle \mathbf x| \normord{\sigma_i \hat{v}_j} \hat{\rho} | \mathbf x \rangle 
	=  \int d \Omega \, \left( 3s_i \frac{p_j}{m} - \mu_B \partial_{p_j} T_i \right) f
\label{eq:spin-moment}
\end{equation}
where \emph{the factor of $3$ is needed only in the first term}.
Also using the definition of $f$, we find 
\begin{equation}
	\int d^2 s \, s_i s_j f = \frac{1}{3}\delta_{ij} \int d^2s \, f 
\end{equation}
and, importantly 
\[
	\int d^2 s \, s_i \partial_{s_j}   f 
	= 
	\frac{1}{4\pi}\int d^2 s \, s_i \Tr [(\sigma_j - s_j s_k \sigma_k) W] = 0
\]
because all odd moments over the sphere vanish by reflection symmetry.

Thus, to find 
\[
	\partial_t \int d \Omega \, 3 \mathbf s f  
	= 3 \int d \Omega \, \mathbf s \partial_t f
\]
we again look at the terms in the kinetic equation.
The Lorentz force term will not contribute anything: the electric force is independent of $\mathbf p$, so we get $\int d^3 p \, \nabla_p f = 0$; when integrating the magnetic force term by parts we get $\nabla_p \cdot (\mathbf p \times \ldots) = 0$.
For the first term, we get 
\begin{multline}
	\int d \Omega \, 3s_i \left( \frac{p_j}{\epsilon} 
	- \mu_B \partial_{p_j} T_l (s_l + \partial_{s_l} )\right) \partial_{x_j} f 
	\\ = 
	\int d \Omega \, \left( 3 s_i \frac{p_j}{\epsilon} 
	- \mu_B \partial_{p_j} T_i \right) \partial_{x_j} f 
\label{eq:spin-first-term}
\end{multline}
using the identities above.

The index structure in the spin torque term will be \[
s_i \epsilon_{jkl} s_k T_l [\sigma_j - s_j s_n \sigma_n]
\]
where the term $4$:th order in $s_i$ does not contribute because it is contracted with the anti-symmetric Levi-Civita.
Thus the spin-torque term, unsurprisingly and reassuringly, gives the spin precession as in the Bargmann-Michel-Telegdi equation~\cite{Bargmann1959}:
\begin{multline}
	\int d^2 s \, s_i \epsilon_{jkl} s_k T_l \partial_{s_j} f =
	\frac{3}{4\pi} \int d^2 s \, s_i \epsilon_{jkl} s_k T_l \Tr[\sigma_j W ] \\
	 = \epsilon_{jil} T_l \Tr [\sigma_j W] 
	 = \left( \mathbf T \times \langle\bm{\sigma}\rangle \right)_i
	 = \left( \mathbf{T} \times \int d^2 s \, 3\mathbf s f \right)_i .
\end{multline}

The term containing $ \nabla_x \tilde T $ in the kinetic equation is treated similarly to the first.
It gives an $x$-derivative on the function in~\cref{eq:spin-moment} complementing that on $f$ in~\cref{eq:spin-first-term}.

In conclusion, after multiplying by $\hbar / 2$, giving the spin angular momentum, we find 
\begin{multline}
	\frac{\hbar}{2} \partial_t \int d^2 s \, 3 s_i f 
	= -\frac{\hbar}{2} \partial_{x_j} \int d^2 s \, \left (s_i \frac{p_j}{\epsilon} - \mu_B \nabla_{p_j} T_i \right)  f  
	\\ - \left( \mathbf{T} \times \int d^2 s \, 3\mathbf s f \right)_i . 
\label{eq:spinTorque}
\end{multline}


\subsection{Angular momentum}

Let $\bm{\Pi} = \mathbf D \times \bB + \langle \mathbf p \rangle$ be the total linear momentum, and define the orbital angular momentum $\mathbf L = \mathbf r \times \bm{\Pi}$.
Then 
\begin{align}
	\partial_t L_i & = \epsilon_{ijk} r_j \partial_t P_k = - \epsilon_{ijk} r_j \partial_n T_{kn} \\
	& = -\partial_n \big( \epsilon_{ijk} r_j T_{kn} \big) + \epsilon_{ijk} T_{kj}
\end{align}
and we see that the orbital angular momentum would be conserved, were $T_{kj}$ symmetric.

The source of orbital angular momentum is $-\epsilon_{ijk} T_{jk}$.
For our stress tensor, this is
\begin{multline}
	\mathbf H \times \mathbf B + \bE \times \mathbf D  - \langle \mathbf{p} \times \mathbf{v} \rangle \\
	= -\mathbf M \times \bB + \bE \times \mathbf P - \langle \mathbf p \times \mathbf v \rangle
\end{multline}
using that $\mathbf H = \mathbf B - \mathbf M, \mathbf D = \bE + \mathbf P$, so only the magnetization/polarization contributes to the cross product.

Now, 
\begin{align}
	\mathbf p \times \mathbf v & = \frac{3\mu_B \mathbf p}{\epsilon(\epsilon +m)} \times (\mathbf E \times \bs) \\
	\bE \times \mathbf P & 
	= \bE \times \left\langle \frac{3\mu_B}{\epsilon(\epsilon + m)} \mathbf p \times \bs \right\rangle
	\end{align}
so 
\begin{multline}
	\bE \times \mathbf P - \langle \mathbf p \times \mathbf v \rangle 
   \\ = \frac{3\mu_B}{\epsilon(\epsilon +m)}\big\langle \bE \times (\mathbf p \times \bs) + \mathbf p \times (\bs \times \bE)  \big\rangle 
	\\
	 =
	\frac{3\mu_B}{\epsilon(\epsilon + m)} \big\langle \bs \times (\mathbf p \times \bE) \big\rangle , 
\end{multline}
using the Jacobi identity for the cross product.

We then recognize that 
\begin{align}
	\partial_t L_i + \partial_n \epsilon_{ijk} r_j T_{kn} 
	= - \int d \Omega \, \left( 3\mathbf s \times \mathbf{T} \right)_i f . 
\end{align}
But this source term is exactly the negative of the spin torque in~\eqref{eq:spinTorque}, so the total angular momentum, orbital plus spin, is conserved.
Hence, the reasons to prefer a symmetric stress-energy tensor, discussed above, are not relevant in our case. 


\section{The thermodynamic background Wigner function}
\label{sec:thermodynamics}

The aim is to find the thermodynamic background Wigner function for electrons in a constant magnetic field $B_0 \mathbf{\hat{z}}$, and also to
compute the associated background magnetization.
We will divide the treatment into the non-relativistic and the relativistic regime.
Before we look into specific cases, we apply spherical coordinates in spin-space (where $\theta_{s}$ is the angle with the $z$-axis) and introduce the division of the equilibrium Wigner function $f_0$ into its spin-up and spin-down parts (see e.g. Ref.~\cite{ZamanianNJP}) according to  
\begin{equation}
 f_0 = \frac{1}{4\pi} F_{0+}(1+\cos \theta_{s}) + \frac{1}{4\pi} F_{0-}(1 - \cos \theta_{s})
 \label{eq:spinsplit}
\end{equation} and \begin{equation}
	n_{0\pm } =\frac{1}{4\pi} \int F_{0\pm }(1 \pm \cos \theta_{s}) \, d\Omega =\frac{1}{4\pi}\int F_{0\pm } \, d\Omega
\end{equation} 
where  $n_{0\pm}$ is the number density of the spin-up and spin-down  populations respectively, and we write  $n_0 = n_{0+} + n_{0-}$.
The factors $(1 \pm \cos\theta_s)$ correspond to the projection operators $1 \pm \sigma_z$.

We note that non-relativistically, the background magnetization $\mathbf{M}_0 = \mu_B \int \bs f_0 (\bp,\bs) \, d\Omega = M_0 \mathbf{\hat{z}}$ is
given by $M_0 =\mu_B n_{0+}-\mu_B n_{0-}=R\mu_B n_0 $ where the
thermodynamic factor $R$ is defined by
\begin{equation}
	 R\equiv \frac{n_{0+}-n_{0-}}{n_0 };	
\end{equation}
i.e., it is a functional of the distribution.
Below, the value of $R$, i.e., the degree of spin-polarization, will be presented for a few specific cases and the relativistic generalization for the background magnetization will be given. 

\subsection{The non-relativistic regime}

For the non-relativistic case the characteristic kinetic energy $E_{k}$ should be much smaller than the electron rest mass energy.
$E_k$ is the thermal energy $k_B T$ or the (non-relativistic) Fermi energy  $E_F = (3\pi ^2 n_0)^{2/3} \hbar^2 /2 m$, whichever is larger.
Moreover, we assume the Zeeman energy $\mu_B B_0$ to fulfill $\mu_B B_0 \ll  E_k$.
If this is fulfilled Landau quantization is not significant, which means that the energy states are continuous to a good approximation.
The general non-relativistic expression for the Wigner function has been computed in Ref.~\cite{ZamanianNJP} (see Eq.~(59) in that work).
In the limit of $\mu_B B_0  \ll E_{k}$, this expression reduces to
\begin{equation}
f_0 (\bp, \bs) = C \sum_\pm
	\frac{\left( 1\pm \cos \theta_{s}\right) }
    {\exp \left[( p^2/2m \mp \mu_B B_0  + \mu_{c})
	/k_B T \right]+1}  \label{Non-relativistic general}
\end{equation}%
where the normalization constant $C$ can be chosen such as to fulfill $\int
f_0 (\bp, \bs) \, d\Omega = n_0$.
The chemical potential $\mu_c$ coincides with the Fermi energy for $T\ll T_F $, and for the opposite ordering it suffices to know that $\mu_{c}$ is large and negative.
For $T\gg T_F$ we can use $\exp( -\mu_c/k_ BT ) \gg 1$ and get 
\begin{widetext}
\begin{equation}
	f_0 (\bp, \bs) = C\exp \left( \frac{p^2}{2mk_B T}\right) 
	\left[
		\exp \left( -\frac{\mu_B B_0}{k_B T}\right) \left( 1+\cos \theta_{s}\right)
		+ \exp \left( +\frac{\mu_B B_0 }{k_B T}\right) \left( 1-\cos
\theta_{s}\right) \right]   \label{eq:non-rel-non-deg}
\end{equation}
\end{widetext}
where the normalization condition gives us 
\begin{equation}
	C = \frac{4\pi n_0 (2\pi k_B T)^{3/2}}{\exp \left( -\frac{\mu_B B_0 }{k_B T}\right) + \exp \left(+\frac{\mu_B B_0 }{k_B T}\right) }
\label{eq:normalization-1}
\end{equation}

For the non-degenerate case given by \eqref{eq:non-rel-non-deg}, it is easily confirmed that the $R$-factor is given by the textbook result \begin{equation}
	 R \big[T \gg T_F, E_k \ll m \big] = \tanh \left( \frac{\mu_B B_0 }{k_B T} \right) .
	 \label{eq:R-factor}
\end{equation}

Next we consider the fully degenerate case where $T = 0$.
As a result, the number densities of spin-up and down particles are given by the volumes of the respective Fermi spheres, limited by the Fermi-momentum $p_{F\pm}=\sqrt{2m (E_F \mp \mu_B B_0)}$ for spin-up and down states respectively.
Thus we have \begin{equation}
R\big[ T \ll T_F, E_k \ll m \big]	
	= \frac{p_{F+}^3 - p_{F-}^3}{p_{F+}^3 + p_{F-}^3}
\approx \frac{3}{2}\frac{\mu_B B_0}{k_B T_F}.
	 \label{R-factor-fd-nr}
\end{equation}%
The approximation made in \eqref{R-factor-fd-nr} is a Taylor expansion to first order in  $\mu _B B_0 /k_B T_F $, in line with the general condition  $\mu_B B_0 \ll E_k $, which is needed to avoid the complications related to strong Landau quantization. 

\subsection{The relativistic regime}

Since $\mu_B B_0 \ll E_k$ still holds relativistically, Landau
quantization is not an issue.
As a result, most of the previous section can be copied by simply replacing the velocity dependence with a momentum-dependence, the non-relativistic kinetic energy with $(\gamma - 1)m$, replacing $\mu_B B_0 $ with $\mu_B B_0 /\gamma $ and now using the relativistic Fermi energy $E_F = \left( (m)^2 + c^4 p_F^2 \right)^{1/2}  -m$ where $p_F^ 2= (3\pi
^2 n_0)^{2/3}\hbar^2 $.
Here we may neglect the corrections of the relativistic factor due to the magnetic dipole energy \cref{eq:velocity} (there is no electric field in equilibrium, so the $\mathbf E \times \bs$ term vanishes) and  use $\gamma = (1 + p^2/m^2 c^2)^{1/2}$.
In the nondegenerate case (large negative chemical potential) we immediately get
\begin{widetext}
\begin{equation}
f_0 (\bp,\bs)=C\exp \left( -\frac{(\gamma -1)m}{k_B T}%
\right) \left[ \exp \left( -\frac{\mu_B B_0 }{\gamma k_B T}\right)
\left( 1+\cos \theta_{s}\right) +\exp \left( +\frac{\mu_B B_0 }{\gamma
	k_B T}\right) \left( 1-\cos \theta_{s}\right) \right]  .
\label{Relativistic non-degerate}
\end{equation}
The corresponding thermodynamic factor thus becomes
\begin{equation}
R\big[T \gg T_F \big] 
  = \frac{
  		\int [p^2\exp \left[ (-(\gamma - 1) m + \mu_B B_0 /\gamma )/k_B T\right]
  		- p^2\exp \left[ (-(\gamma - 1)m + \mu_B B_0 /\gamma )/k_B T\right] dp}
  	{
  		\int [p^2\left[ (-(\gamma - 1)m + \mu_B B_0 /\gamma )/k_B T\right]
  		+ p^2\exp \left[ (-(\gamma - 1)m +\mu_B B_0 /\gamma )/k_B T\right] dp} .
\label{eq:R-factor-3}
\end{equation}
Since $\mu_B B_0 $ is a correction term, we may Taylor expand the exponentials, in which case we get
\begin{equation}
R\big[ T\gg T_F] 
  =\\  \frac{\mu_B B_0 }{k_B T}
	\frac{\int \frac{p^2}{\gamma } \exp \left[ -(\gamma -1)m/k_B T\right] \, dp}
	{\int p^2\exp \left[ -(\gamma -1)m/k_B T\right] \, dp}  \label{eq:R-factor-4}.
\end{equation}
which agrees with the non-relativistic expression in the limit of $(\gamma -1) \ll 1 $.

For a fully degenerate relativistic system we again use $f_0 = F_{0+}(1+\cos \theta_{s}) + F_{0-}(1 - \cos \theta_{s})$ but with 
\begin{equation} 
	F_{0\pm} = \begin{cases} 1/(2\pi \hbar )^3 & (\gamma - 1)m\mp \frac{\mu_B B_0 }{\gamma }\leq E_F  \\
0 & \text{otherwise} 
\end{cases}. \label{eq:Energy-rel}
\end{equation}

Thus the $R$-factor is determined by
\begin{equation}
R\big[T \ll T_F \big] 
  =\frac{p_{F+}^3 - p_{F-}^3}{p_{F+}^3 + p_{F-}^3}
\label{eq:R-factor-5}
\end{equation}%
where $p_{F\pm }$ is determined from 
    \begin{equation}
        \sqrt{1 + \frac{p_{F\pm}^2}{m^2c^2}} - 1 = \frac{E_F \pm \mu_B B_0 / \gamma_F}{m} \\ \simeq \frac{E_F }{m} \pm \frac{\mu_B B_0 }{m + E_F}  \label{eq:Fermi-momentum}
    \end{equation}
where we have used that the relativistic factor near the Fermi surface fulfills $\gamma_F  \simeq 1+E_F /m$.
\end{widetext}
Simplifying \cref{eq:R-factor-5} using $\mu_B B_0 \ll m$ and $\mu_B B_0 \ll E_F $ results in the final expression
\begin{equation}
R\big[T \ll T_F]	
	= \frac{3\mu_B B_0 }{\left( \left( E_F /m+1\right) ^2-1\right) m} 
	 \label{eq:R-factor-6}
\end{equation}
which coincides with the non-relativistic expression in the limit $E_F \ll m$.

It should be noted that the expression $M_0 =R\mu_B n_0 $ requires that each particle contributes with $\pm \mu_B $ to the magnetic moment.
Since this is only true in the rest frame, we must compensate. 
If the magnetic dipole energy is large this would be complicated, but for $\mu_B B_0 \ll m$ a simplified calculation suffices.
The result is that the magnetization is reduced in proportion to the gamma-factor, i.e.,
\begin{equation}
	M_0 = R\mu_B n_0 \left\langle \frac{1}{\gamma}\right\rangle
    = R \mu_B n_0 \frac{ \int \frac{1}{\gamma} f_{0p} \, d^3p}{\int f_{0p} \, d^{3}p}
    \label{eq:Magnetization-relativistic}
\end{equation}
where $f_{0p}$ is the reduced momentum distribution function that would apply in the absence of spin-polarization (i.e. a relativistic Fermi-Dirac or a Synge-Jüttner distribution for the two cases considered above).
The above expression gives the correct magnetization to first order in an expansion $\mu_B B_0 /m$.
It should be stressed that the expression for $M_0 $ becomes significantly more complicated if higher order contributions are required.

Physically we can note that relativistic effects decrease the total background magnetization in two ways.
Firstly, for relativistic particles the energy difference between the spin-up and spin-down states (in the laboratory frame) is smaller, reducing the difference in number density between the spin-up and down states (see e.g.~\cref{Relativistic	non-degerate}).
Secondly, the contribution to the magnetic dipole moment in the laboratory frame is smaller for a moving particle, reducing the magnetization a second time (see e.g.~\cref{eq:Magnetization-relativistic}).


\section{Dispersion relation for linear waves}

\label{sec:disprel}

The enhanced complexity of \cref{eq:evolution}, compared to non-relativistic theories, introduces some technical obstacles already in linearized theory. 
Our purpose here is to address these difficulties, and present a general calculation method. 

\begin{widetext}
After linearization \cref{eq:evolution} can be written as
	\begin{multline}\label{linerized vlasov eq.}
	\frac{\partial f_1}{\partial t} + \frac{\bp}{m}\cdot \nabla_x f_1 
	- \mu_B m \Big[
		\bB_0 \cdot (\bs +\nabla_s) 
	\Big]  \nabla_p \Big( \eps \Big)\cdot \nabla_x f_1
	+ \frac{q}{m}(\bp \times \bB_0) \cdot \nabla_p f_1 \\
	-q\mu_B m \Big [
		\bB_0\cdot (\bs + \nabla_s)
	\Big] \Big( \nabla_p\eps \times \bB_0 \Big) \cdot \nabla_pf_1
	+ \frac{2\mu_B}{\hbar}(\bs \times \bB_0) \cdot \nabla_s f_1 \\
	= -q\bE \cdot \nabla_p f_0 - \frac{q}{m} \Big[
		\bp \times \bB_1
	\Big] \cdot \nabla_p f_0
	+q\mu_B m\left[
		\Big( \bB_1 -\frac{\bp \times \bE}{2m} \Big)\cdot (\bs +\nabla_s)
	\right] \Big (\nabla_p \eps \times \bB_0 \Big ) \cdot\nabla_p f_0 \\
	+q\mu_B m \Big[
		\bB_{0}\cdot \, (\bs +\nabla_s)
	\Big ] \Big(\nabla_p \eps \times \bB_1 \Big) \cdot \nabla_p f_0
	- q\mu_B \left\{
		\nabla_p\left [
			\frac{\bp \times \bE}{2m}\cdot(\bs + \nabla_s)
		\right ]\times \bB_0
	\right\}\cdot \nabla_pf_0\\
	-\mu_B\nabla_x \Big [
		\bB_1\cdot(\bs + \nabla_s)
	\Big ] \cdot \nabla_pf_0 
	+\frac{\mu_B}{2m}\nabla_x\Big[
		(\bp \times \bE) \cdot (\bs +\nabla_s)
	\Big]\cdot \nabla_pf_0
	-\frac{2\mu_B}{\hbar}\left [
		\bs \times\Big(\bB_1-\frac{\bp \times \bE}{2m} \Big)
	\right ]\cdot \nabla_sf_0.
\end{multline} 
Here $\theta_s$ is the angle between the spin and $\mathbf{B}_0$.
Some of the difficulties in \cref{linerized vlasov eq.}, as compared to the classical Vlasov equation follow from $ \gamma = \epsilon / m $ being a function of momentum.
However, since it is well-known from classical relativistic theory theory how to deal with this, we will focus on some of the other subtleties, and use the lowest order approximations, $ \epsilon / m \approx 1$,  $ \nabla_p ( 1 / \epsilon ) \approx -2 \bp / m^3 $.  

From now on we consider a restricted geometry, namely transverse waves with
\begin{equation*}
	\mathbf{k} = k_{\perp}\be_x \quad
	\bE_1  = E_1 \be_z \quad
	\bB_1 = B_1 \be_y ,
\end{equation*}
where $\mathbf{k}$ is the wave vector.
The perturbed quantities follow the plane-wave ansatz according to $f_1 = \Tilde{f}_1e^{i(\mathbf{k} \cdot \mathbf{x}-\omega t)}$.
We express the momentum of the particles in cylindrical coordinates $(p_{\perp}, \varphi_p, p_z)$ and the spin in spherical coordinates, $\bs = \sin\theta_s \cos\varphi_s \be_x+ \sin\theta_s \sin\varphi_s \be_y +  \cos\theta_s\be_z$.

Furthermore we make an expansion of $\Tilde{f}_1$ in eigenfunctions of the operators of the right-hand side of~\cref{linerized vlasov eq.}~\cite{lundin2010linearized}
\begin{equation}
	\label{Eigen function}
	\widetilde{f_1}(\bp,\bs) =
	\sum_{\alpha,\beta} g_{\alpha\beta}(p_{\perp}, p_z, \theta_s)
	\psi_{\beta}(\varphi_p, p_{\perp})
	\frac{e^{i\alpha\varphi_s}}{\sqrt{2\pi}} ,
\end{equation}
where
\begin{align}
	\label{Bessel eigen function}
	\psi_{\beta}(\varphi_p, p_{\perp})
	= \frac{1}{\sqrt{2\pi}} \exp\Bigg[
		i\Big(
			\beta\varphi_p - \frac{k_{\perp}p_{\perp}}{m\omega_{ce}}  \sin\varphi_p
		\Big)
	\Bigg] 
	= \frac{1}{\sqrt[]{2\pi}}
	\sum_{\tau}
		J_\tau \Big( \frac{k_{\perp} p_{\perp}}{m\omega_{ce}} \Big)
		e^{i(\beta'-\tau)\varphi_p},
\end{align}
where $J_\tau$ is the Bessel function of first kind and $\omega_{ce} = qB_0/m$ is the electron cyclotron frequency.
Here, Greek summation indices take integer values from $-\infty$ to $+\infty$ and we will suppress the argument of the Bessel functions, as it is always $k_\perp p_\perp / m \omega_{ce}$.
Only $\alpha = \pm 1$ will contribute in \cref{Eigen function}.
From now on, we will additionally use Latin summation indices that take only the values $\pm 1$ to distinguish between the two types of sums. 

Using this eigenfunction expansion together with
\begin{subequations} \label{simplification}
	\begin{align}
	\frac{q}{m}(\bp \times \bB_0) \cdot \nabla_pf_1 & = 
	\omega_{ce}\frac{\partial f_1}{\partial \varphi_p} \\
	\frac{2\mu_B m}{m\hbar} (\bs \times \bB_0) \cdot \nabla_sf_1 & =
	-\omega_{cg}\frac{\partial f_1}{\partial\varphi_s}.
	\end{align}
\end{subequations}
where $\omega_{cg} = (g/2)\omega_{ce}$ in~\cref{linerized vlasov eq.}, we get
	\begin{multline}\label{LHS of linerized vlasov transvers}
	\Bigg \{
		-i\omega + \frac{ip_{\perp}k_{\perp}}{m} \cos\varphi_p \left [
			1+ \frac{\mu_B B_0}{m}
			(\cos\theta_s - \sin\theta_s \frac{\partial}{\partial\theta_s})
		\right]
		-\omega_{ce}\left[
			1+ \frac{\mu_B B_0}{m}
			(\cos\theta_s - \sin\theta_s \frac{\partial}{\partial\theta_s}) 
			\right] \frac{\partial}{\partial \varphi_p}
		-\omega_{cg}\frac{\partial}{\partial \varphi_s}
	\Bigg\} f_1
\\
	= \textnormal{RHS},
	\end{multline}
where $\textnormal{RHS}$ is the right-hand side of~\cref{linerized vlasov eq.}, simplified according to the above assumptions.

The differential equation~\cref{LHS of linerized vlasov transvers} is relatively hard to solve analytically.
However the troublesome operator $\cos \theta_s - \sin\theta_s \partial_{\theta_s}$ appears with a factor $\mu_B B_0/m$, which is of order of unity for magnetars, but is much smaller for other known environments~\cite{Harding2006}.
Hence we can solve this equation for $f_1$ by using perturbation theory.
Expanding $f_1 = f_{10} + f_{11}$, where $f_{11}$ is first order in $\mu_B B_0/m$, $f_{10}$ is the solution to~\cref{LHS of linerized vlasov transvers}, keeping only zeroth order terms in $\mu_B B_0/m$ in the right-hand side:
\begin{equation}
	f_{10} = \sum_{n, \rho,\beta} J_{\beta} e^{i\varphi_p(\beta-\rho)}  
	\Big[
		A_{\rho} + \Big(B_{\rho}+ C_{\rho} \Big)e^{ in\varphi_s}
	\Big].
\end{equation}
The first order term $f_{11}$ is then obtained by taking $f_1 = f_{10}$ in all first-order terms in~\cref{LHS of linerized vlasov transvers}:
	\begin{equation}
	f_{11} = \frac{\mu_B B_0 \omega_{ce}}{m}\Big(\cos\theta_s - \sin\theta_s \frac{\partial}{\partial \theta_S}\Big)
	\sum_{\tau,\beta,\alpha,\rho}
	e^{i\varphi_p(\beta-\rho)} \rho
	J_{\beta}J_{\alpha}J_{\rho + \alpha -\tau}
	\sum_{n }\Bigg[
		\frac{A_{\tau}}{\omega-\omega_{ce}\rho}
		+ \frac{B_{\tau} + C_{\tau} }{\omega-\omega_{ce}\rho + n\omega_{cg}}e^{ in \varphi_s}
	\Bigg].
	\end{equation}
In these expansions,
\begin{align}
    A_{\rho} & = -iqE\frac{\partial f_0}{\partial p_z} \frac{J_{\rho}}{\omega -\omega_{ce}\rho} \\
    B_{\rho} & = -i\Big(
		\sin\theta_s + \cos\theta_s \frac{\partial}{\partial \theta_s}
	\Big)
	\frac{\partial f_0}{\partial p_{\perp}}
	\sum_{n}
	\frac{\mu_B}{\omega - \omega_{ce} \rho + n\omega_{cg}}
	\Bigg[
		\frac{qE B_0}{4m}J_{\rho -n} 
		+ n \frac{ B_1 m \omega_{ce}}{2p_{\perp}} \rho J_{\rho}
		+\frac{E k_{\perp} p_{\perp}}{4m} J'_{\rho -n}
	\Bigg] \\
    C_{\rho} & = \frac{\partial f_0}{\partial \theta_s}
	\sum_{n} \frac{i\mu_B / \hbar }{\omega - \omega_{ce}  + n\omega_{cg}}  \Bigg[
		B_1 J_{\rho}
		+\frac{Ep_{\perp}}{2m} J_{\rho  -n}
	\Bigg].
\end{align}

Now with $f_1$ expressed in terms of $f_0$, we can split $f_0$ into its spin parts, using the notation $f_{0\pm}=\frac{1}{4\pi} F_{0\pm}(1\pm \cos{\theta_s})$ as in the previous section, \cref{eq:spinsplit}.
This will give additional sums over $\pm$, analogous to sums over the Latin indices already used.
In determining the dispersion relation, we calculate the total current density $\mathbf{J} = \mathbf{j}_f  + \frac{\partial \mathbf P}{\partial t} + \nabla \times \mathbf{M}$ (see~\cref{eq:j-free,eq:polarization,eq:magnetization}).
Using Maxwell's fourth equation, \cref{eq:ampere}, the dispersion relation is given by
\begin{equation} \label{Dispersion relation general}
    \omega^2 = k^2  + \frac{\pi^2q^2\omega }{m}\sum_{\beta}
    \int d^2p \, J_{\beta} \left[ 
    	\chi^f_\beta + \chi^M_\beta + \chi^P_\beta 
    \right],
\end{equation}
where $\chi^f_\beta$ is the contribution from the free current density
\begin{equation}
	\chi^f_\beta = \sum_{\pm} \frac{8J_{\beta}}{\omega-\omega_{ce}\beta}
 	\Bigg[
 		1 \pm \frac{ \mu_BB_0}{m}  \pm \frac{ \mu_BB_0 \omega_{ce}\beta }{m}\sum_{\alpha,\tau }
 		\frac{J_{\alpha} J_{\alpha + \beta - \tau}}{\omega - \omega_{ce}\tau}
	\Bigg]F_{0\pm},   
\end{equation}
$\chi^M_\beta$ is the contribution from the magnetization current density
\begin{equation}
    \chi^M_\beta =\sum_{\pm,n }  \frac{n\hbar^2k /m }{\omega- \omega_{ce}\beta  + n \omega_{cg}} 
    \Bigg[ 
	    \Bigg(
	    	\frac{qB_0}{2m} J_{\beta -n} - \frac{nm \omega_{ce}k \beta}{p_{\perp}\omega} J_{\beta} 
    	+  \frac{kp_{\perp}}{2m} J'_{\beta - n }
    	\Bigg) \frac{\partial F_{0\pm}}{\partial p_{\perp}} 
    	\mp  \frac{1}{\hbar} \Bigg(
    		\frac{2k}{\omega}J_{\beta}- \frac{p_{\perp}}{m} J_{\beta -n}
    	\Bigg) F_{0\pm}
    \Bigg],
\end{equation}
and $\chi^P_\beta $ is the contribution from the polarization current density
\begin{equation}
    \chi^P_\beta = \sum_{\pm, n}  \frac{-n \hbar^2\omega p_{\perp} / 2m^2 }{\omega - \omega_{ce}(\beta+n)+  n \omega_{cg}} 
    \Bigg[ 
	    \Bigg(
	    	\frac{qB_0}{2m} J_{\beta} - \frac{nm \omega_{ce}k (\beta +n)}{p_{\perp}\omega}J_{\beta +n} 
    		+  \frac{kp_{\perp}}{2m}J'_{\beta}
	    \Bigg) \frac{\partial F_{0\pm}}{\partial p_{\perp}} 
	    \mp \frac{1}{\hbar} \Bigg(
			\frac{2k}{\omega}J_{\beta + n}- \frac{p_{\perp}}{m} J_{\beta}
    	\Bigg) F_{0\pm}
    \Bigg].
\end{equation}
\end{widetext}
Note that only in $\chi^f_\beta$ do we have contributions from $f_{11}$, this is because both $B_{\rho}$ and $C_{\rho}$ vanish in $f_{11}$ when using $f_{0\pm}=\frac{1}{4\pi}F_{0\pm}(1\pm \cos{\theta_s})$.

\cref{Dispersion relation general} contains both classical and quantum modes.
Since we are interested in the effects of the spin on the dispersion relation, we consider a regime where the spin effects are comparable to the classical ones.
Considering 
\begin{equation}
	\omega \approx 	\Delta\omega_{ce} := \omega_{cg} - \omega_{ce},
\end{equation}
in this regime the denominators $\omega - \omega_{ce}\beta $, $\,\omega- \omega_{ce}\beta \pm \omega_{cg}$ and $\omega- \omega_{ce}\beta \pm  \Delta \omega_{ce}$ are minimized for $\beta= 0, \pm 1$,  and $0$ respectively.
Keeping only these values of $\beta$ in the summation over $\beta$ and focusing on the short Larmor radius regime, i.e., $kp_{\perp}/m\omega_{ce} \ll 1$,  we expand the Bessel functions to second order.
We refer to \cref{Harmonic limit} for more details.

Furthermore, to compute the integrals in \cref{Dispersion relation general} explicitly, we must specify the background distribution function. 
Considering a Fermi-Dirac distribution for low temperatures $T \ll T_F$, where $T_F$ is the Fermi-temperature, the dispersion relation in equation~\cref{Dispersion relation general} is 
\begin{equation} \label{Dispersion relation specific}
    \omega(k) = \Delta\omega_{ce} \Bigg[ 
   	 1 - \frac{7\hbar^2 \omega_{p}^2k^2}{20m^2(\Delta\omega_{ce}^2-k^2-\omega_p^2 )}
		\bigg(
			1 + \frac{k^2}{2\delta \omega_{ce} }
    	\bigg)
    \Bigg], 
\end{equation}
where $\delta = g/2 - 1$. For $k \rightarrow 0$, $\omega \rightarrow \Delta\omega_{ce}$ as expected.
The equation above shows that $\omega$ diverges from $\Delta \omega_{ce}$ for larger wave numbers.
This divergence is proportional to the square of the Compton wavelength, but since the solution is in the short Larmor radius regime, $k$ should satisfy $k \ll m \omega_{ce}/p_{\perp}$. 

\section{Discussion} 
\label{sec:discussion}

The present quantum kinetic relativistic model for electrons, first derived in Ref.~\cite{PhysRevE.96.023207}, is based on the Dirac equation, where a phenomenological adjustment of the spin $g$-factor has been made to account for the QED-contribution.
It generalizes the relativistic Vlasov equation to include the spin dynamics, while still allowing for a fully relativistic motion.
In comparison with the models found in Ref.~\cite{ZamanianNJP} it is extended to contain spin-orbit interaction, a contribution to the polarization current associated with the spin, and Thomas precession.
As compared to the model of Ref.~\cite{asenjo2012semi} it allows for fully relativistic motion (i.e., relativistic gamma factors not close to unity).
Here we have continued studying the model, including the full set of conservation laws for energy, momentum and angular momentum, as well as an expression for the stress-energy tensor.
Moreover, the thermodynamic background Wigner function and the background magnetization are given for the degenerate and non-degenerate cases, in both the relativistic and non-relativistic cases.

A complication in the present theory when it comes to practical calculations is the non-trivial relation between the momentum variable and the velocity, involving the spin state.
To illustrate how this can be handled analytically, treating the spin-dependence perturbatively, we have calculated the linear dispersion relation for the case of perpendicular propagation across an external magnetic field.
Finally a simple limiting case with $\omega \approx \Delta \omega _{ce}$ has been presented. 

\sisetup{range-phrase=-, range-units=single}
Before we go on with the technical aspects, let us first discuss some specific systems where the current model is of particular interest, i.e.  plasmas where the particle motion is relativistic and the electrons are spin-polarized.
As described in~\cref{sec:thermodynamics}, the thermodynamic background distribution has a degree of spin polarization of the order $\mu_B B_0/k_BT$ for a non-degenerate system and of the order $\mu_B B_0 /k_B T_F$ for degenerate systems.
For astrophysical objects such as pulsars ($B \sim \SI{e8}{T}$ at the surface) and magnetars ($B \sim \SIrange{e10}{e11}{T}$ at the surface) \cite{uzdensky2014extreme,Harding2006}, we typically have non-relativistic temperatures $T \sim \SI{e6}{K}$ in the atmosphere~\cite{page1996temperature}.
This corresponds to a large electron spin-polarization $0.1 < \mu_B B_0 /k_B T < 1$ even at a considerable distance from the neutron star surface.
While the electron temperature may not be of relativistic magnitude, the particle interactions with the pulsar fields nevertheless induce relativistic motion in the atmosphere~\cite{mahajan2013ultra,daCosta1991relativistic}.
For electrons at the pulsar or magnetar surface, on the other hand, the degeneracy makes the particles strongly relativistic with a Fermi temperature of the order $T_F \sim \SI{e12}{K}$ \cite{lattimer2004physics}, which, depending on the magnetic field, corresponds to a degree of spin polarization $0.0001 < \mu_B B_0 /k_B T_F < 0.1$.

Thus for electrons belonging to the neutron star surface, we note that spin polarization is pronounced for magnetars but not as much for pulsars.
Turning to laboratory applications, laser-plasma interactions with dense targets are of particular interest~\cite{shi2018laser,lindman2010laser,wagner2004laboratory,PhysRevA.96.043407,delSorbo2018polarization,PhysRevLett.122.154801,marklund2006nonlinear,DiPiazza2012,PhysRevE.83.036410,PhysRevLett.105.105004}.
While such systems are not initially spin-polarized, theory aided by particle-in-cell simulations have predicted~\cite{shi2018laser}  that quasi-static magnetic field with a field strength of the order $B_0 \sim \SI{e5}{T}$ can be formed.
It should be stressed that these estimates are fully consistent with experiments, see e.g. Refs~\cite{lindman2010laser,wagner2004laboratory}.
Before the final stage of laser-plasma compression, where signficant electron heating may occur, this corresponds to a spin polarization of the order $0.01 < \mu_B B_0 /k_B T_F  < 0.1$, at the same time as the electrons are relativistic due to the quiver motion in the laser field \cite{marklund2006nonlinear,DiPiazza2012}.

Here we have mainly been concerned with systems that, besides being relativistic, have a thermodynamic background distribution that is spin polarized.
However, it is worth noting that interaction with a dynamical electromagnetic fields by itself can lead to spin polarization, see e.g. Refs~\cite{PhysRevA.96.043407,delSorbo2018polarization,PhysRevLett.122.154801,PhysRevE.83.036410,PhysRevLett.105.105004}.
Moreover, we stress that the current theory deviates from simpler models to some extent already for modest gamma factors, see the discussion of hidden momentum in Paper~I.

The current model is of particular interest for strong fields. However,  classical or QED corrected radiation reaction~\cite{nikishov1964quantumI,nikishov1964quantumII,harvey2012radiation} is not covered and neither is strong field vacuum polarization due to QED included.
Thus it is of interest to establish the regime of validity for the model, i.e. to find the conditions when the spin contributions are the most important extensions of the classical physics.
First it should be noted that the non-relativistic spin-contributions are not necessarily dependent on strong fields, in particular the ratio of the magnetic dipole force over the Lorentz force scales as $\hbar k^{2}/m\omega $ for a plane wave field.
Thus, in case we have a short-wavelength plasma perturbation with a low-frequency, the magnetic dipole force should be included independent of the field strength.
While effects of this nature have been covered already in previous non-relativistic works~\cite{ZamanianNJP,hurst2014semiclassical} or weakly relativistic theories~\cite{asenjo2012semi,stefan2013linear}, the theory presented here is needed in case the thermodynamic temperature or the Fermi temperature is relativistic.
In the absence of relativistic temperatures, however, the strong field effects of the present theory, whose relative importance is proportional to $\mu_B B/m$,  need to be compared with the effects of radiation reaction and vacuum polarization. 

The contribution from classical radiation reaction relative to the Lorentz force scales as the dimensionless parameter $R = 2a_{0}^{2}\gamma \alpha \omega /3m$ where $a_0 = qE_0/(\omega m)$ is the normalized vector potential, $E_0$ being the peak field strength and $\omega$ the laser frequency; and $\alpha$ is the fine-structure constant.
Thus the relative importance of radiation reaction in relation to relativistic spin effects, in fields of strength $E$ and $B$, is given by
\begin{equation}
	\frac{mR}{\mu_B B} = \frac{2a_0^2 \gamma \alpha \omega}{3\mu_B B}
	=  \frac{4E a_{0} \gamma \alpha }{3B} .
	\label{eq:comp-radiation reaction}
\end{equation}
For definiteness, if the relativistic particle velocities are induced by strong laser fields, such that $E/B\sim 1$ and  $\gamma \sim a_{0}$, the condition for the spin effects to dominate over radiation reaction becomes
\begin{equation}
	\gamma < \left( \frac{3}{4\alpha} \right)^{1/2} \simeq 10.
	\label{eq:comp-radiation reaction-II}
\end{equation}
Thus for moderately relativistic fields, the relativistic contribution to the spin dynamics is more important than radiation reaction, whereas for strong enough laser fields, the ordering is the opposite.
However, for other types of field-configurations (i.e. not due to lasers), in particular in the vicinity of strongly magnetized objects (e.g. pulsars), we may have $E/B\ll 1$.
As seen from \cref{eq:comp-radiation reaction}, in this case the spin-relativistic effects may dominate over radiation reaction even for $\gamma \gg 10$.

Next, we note that our treatment assumes $\mu_B B / m < 1$, which corresponds to a field magnitude below the critical field strength, i.e. the value for which pair production is exponentially suppressed.
In case pair production does not take place, vacuum polarization scales as $(\alpha /90\pi )E^{2}/E_{\mathrm{cr}}^{2}$ \cite{dittrich2000probing,marklund2006nonlinear}.
Thus, provided the plasma is not dilute~\footnote{
	Particle contributions enter a dispersion relation proportional to a factor $\omega_p^2/ \omega^2$, and hence are suppressed for low densities or high frequencies.
	}
, vacuum polarization is smaller than the spin-relativistic effects by a factor of the order \begin{equation}
	\frac{\alpha}{90\pi} \frac{E}{E_{\mathrm{cr}}}  \label{eq:comp-vacuum-polarization}	
\end{equation}
where we have put $E = B$ to simplify the expression, which is justified since electric and magnetic fields induce effects of the same order.      

Due to the complexity of the kinetic model, it is rather difficult to analyse for problems beyond linear waves in homogeneous media.
Still, due to the richness of the physics included, the present theory can be useful in many different contexts.
Firstly, \crefrange{eq:evolution}{eq:magnetization} can serve as a starting point for deriving simpler models, using e.g. a moment expansion~\cite{haas2010fluid,[{}] [{, especially Paper~IV.}]stefan2014models} of the kinetic evolution equation.
Secondly, the present theory can be used to determine the region of validity of many simpler models.
Thirdly, potentially there are new phenomena to be found from \crefrange{eq:evolution}{eq:magnetization} even for the special case of linear homogeneous plasmas, as we have only covered a few of those possibilities here.
Finally, the mathematical structure of the present theory can help shedding light over long-standing problems such as the Abraham-Minkowski controversy~\footnote{Discussed in Paper I~\cite{PhysRevE.96.023207}, see also references therein.}.  

In this paper we have written the model with the spin as an independent variable, which is convenient for analytical calculations.
One can instead not take the spin transform and formulate the model in terms of the four independent components of the Hermitian matrix $W_{\alpha\beta}$, analogous to Ref.~\cite{hurst2014semiclassical}.
In this case, the computational cost of simulating the model would be that of simulating four models for spin-less particles, plus some extra cost due to additional terms, which is not prohibitive compared to simulating the Vlasov equation.

\begin{acknowledgements}
All authors would like to acknowledge financial support by the Swedish Research Council, grant number 2016-03806.
G.~Brodin, H.~Al-Naseri and J.~Zamanian would also like to acknowledge support by the Knut and Alice Wallenberg Foundation through the PLIONA project.
\end{acknowledgements}

\appendix

\section{The Harmonic Limit}
\label{Harmonic limit}
In obtaining the harmonic dispersion relation in \cref{Dispersion relation specific} from \cref{Dispersion relation general}, several approximations have been done.
Working with $\omega \approx \Delta \omega_{ce}$, gives us possibilities for minimizing the values of the denominators in \cref{Dispersion relation general} for certain values of $\beta$.
Considering only the leading terms where the denominators are minimized according to
\begin{align*}
\chi_f &: \omega -\omega_{ce}\beta \rightarrow \omega\\
\chi_M &: \omega-\omega_{ce}\beta \pm \omega_{ce } \rightarrow \omega \pm \Delta\omega_{ce}\\
\chi_P &: \omega -\omega_{ce} \pm \Delta\omega_{ce} \rightarrow \omega \pm \Delta \omega_{ce},
\end{align*}
where we set $\beta= 0$, $\pm 1$ and $0$ respectively.
Note that when using $\beta=0$ in $\chi_f$ the contribution from $f_{11}$ in~\cref{Dispersion relation general} becomes zero. 

Considering the short Larmor radius regime where the argument of the Bessel functions is small, i.e., $kp_{\perp}/m \omega_{ce} \ll 1$, we can make a second order Taylor expansion of the Bessel functions.
Furthermore, we consider the case where the background distribution is Fermi-Dirac with the low temperature $T \ll T_F$.
Hence we have, $F_{0\pm} 4 \pi p_{F\pm}^3 / 3 \approx  n_0 $.
\cref{Dispersion relation general} is now

\begin{multline}
	\label{Dispersion relation 2}
\Big(\omega^2-k^2-\omega_{p}^2 \Big)
\Big(
\omega^2- \Delta\omega_{ce}^2
\Big)
=-\frac{7\hbar^2\omega_{p}^2k^2 }{10m^2}\bigg(
\omega^2+ \frac{k^2 \delta}{2}
\bigg).
\end{multline}
Since we are looking for $\omega \approx \Delta\omega_{ce}$, we consider the $\omega^2 - \Delta\omega_{ce}^2$-root and approximate $\omega^2$ in the right hand side of~\cref{Dispersion relation 2} to $\Delta\omega_{ce}^2$.
We have finally the dispersion relation in~\cref{Dispersion relation specific}.

\bibliography{Relativistic_applications}{}

\begin{thebibliography}{65}%
\makeatletter
\providecommand \@ifxundefined [1]{%
 \@ifx{#1\undefined}
}%
\providecommand \@ifnum [1]{%
 \ifnum #1\expandafter \@firstoftwo
 \else \expandafter \@secondoftwo
 \fi
}%
\providecommand \@ifx [1]{%
 \ifx #1\expandafter \@firstoftwo
 \else \expandafter \@secondoftwo
 \fi
}%
\providecommand \natexlab [1]{#1}%
\providecommand \enquote  [1]{``#1''}%
\providecommand \bibnamefont  [1]{#1}%
\providecommand \bibfnamefont [1]{#1}%
\providecommand \citenamefont [1]{#1}%
\providecommand \href@noop [0]{\@secondoftwo}%
\providecommand \href [0]{\begingroup \@sanitize@url \@href}%
\providecommand \@href[1]{\@@startlink{#1}\@@href}%
\providecommand \@@href[1]{\endgroup#1\@@endlink}%
\providecommand \@sanitize@url [0]{\catcode `\\12\catcode `\$12\catcode
  `\&12\catcode `\#12\catcode `\^12\catcode `\_12\catcode `\%12\relax}%
\providecommand \@@startlink[1]{}%
\providecommand \@@endlink[0]{}%
\providecommand \url  [0]{\begingroup\@sanitize@url \@url }%
\providecommand \@url [1]{\endgroup\@href {#1}{\urlprefix }}%
\providecommand \urlprefix  [0]{URL }%
\providecommand \Eprint [0]{\href }%
\providecommand \doibase [0]{http://dx.doi.org/}%
\providecommand \selectlanguage [0]{\@gobble}%
\providecommand \bibinfo  [0]{\@secondoftwo}%
\providecommand \bibfield  [0]{\@secondoftwo}%
\providecommand \translation [1]{[#1]}%
\providecommand \BibitemOpen [0]{}%
\providecommand \bibitemStop [0]{}%
\providecommand \bibitemNoStop [0]{.\EOS\space}%
\providecommand \EOS [0]{\spacefactor3000\relax}%
\providecommand \BibitemShut  [1]{\csname bibitem#1\endcsname}%
\let\auto@bib@innerbib\@empty
\bibitem [{\citenamefont {Hu}\ and\ \citenamefont {Keitel}(1999)}]{Hu1999}%
  \BibitemOpen
  \bibfield  {author} {\bibinfo {author} {\bibfnamefont {S.~X.}\ \bibnamefont
  {Hu}}\ and\ \bibinfo {author} {\bibfnamefont {C.~H.}\ \bibnamefont
  {Keitel}},\ }\href {\doibase 10.1103/PhysRevLett.83.4709} {\bibfield
  {journal} {\bibinfo  {journal} {Phys. Rev. Lett.}\ }\textbf {\bibinfo
  {volume} {83}},\ \bibinfo {pages} {4709} (\bibinfo {year}
  {1999})}\BibitemShut {NoStop}%
\bibitem [{\citenamefont {Walser}\ and\ \citenamefont
  {Keitel}(2000)}]{Walser2000}%
  \BibitemOpen
  \bibfield  {author} {\bibinfo {author} {\bibfnamefont {M.~W.}\ \bibnamefont
  {Walser}}\ and\ \bibinfo {author} {\bibfnamefont {C.~H.}\ \bibnamefont
  {Keitel}},\ }\href {\doibase 10.1088/0953-4075/33/6/105} {\bibfield
  {journal} {\bibinfo  {journal} {J. Phys. B At. Mol. Opt. Phys.}\ }\textbf
  {\bibinfo {volume} {33}},\ \bibinfo {pages} {L221} (\bibinfo {year}
  {2000})}\BibitemShut {NoStop}%
\bibitem [{\citenamefont {Walser}\ \emph {et~al.}(2002)\citenamefont {Walser},
  \citenamefont {Urbach}, \citenamefont {Hatsagortsyan}, \citenamefont {Hu},\
  and\ \citenamefont {Keitel}}]{Walser2002}%
  \BibitemOpen
  \bibfield  {author} {\bibinfo {author} {\bibfnamefont {M.~W.}\ \bibnamefont
  {Walser}}, \bibinfo {author} {\bibfnamefont {D.~J.}\ \bibnamefont {Urbach}},
  \bibinfo {author} {\bibfnamefont {K.~Z.}\ \bibnamefont {Hatsagortsyan}},
  \bibinfo {author} {\bibfnamefont {S.~X.}\ \bibnamefont {Hu}}, \ and\ \bibinfo
  {author} {\bibfnamefont {C.~H.}\ \bibnamefont {Keitel}},\ }\href {\doibase
  10.1103/PhysRevA.65.043410} {\bibfield  {journal} {\bibinfo  {journal} {Phys.
  Rev. A}\ }\textbf {\bibinfo {volume} {65}},\ \bibinfo {pages} {043410}
  (\bibinfo {year} {2002})}\BibitemShut {NoStop}%
\bibitem [{\citenamefont {Wen}\ \emph {et~al.}(2016)\citenamefont {Wen},
  \citenamefont {Bauke},\ and\ \citenamefont {Keitel}}]{Wen2016}%
  \BibitemOpen
  \bibfield  {author} {\bibinfo {author} {\bibfnamefont {M.}~\bibnamefont
  {Wen}}, \bibinfo {author} {\bibfnamefont {H.}~\bibnamefont {Bauke}}, \ and\
  \bibinfo {author} {\bibfnamefont {C.~H.}\ \bibnamefont {Keitel}},\ }\href
  {\doibase 10.1038/srep31624} {\bibfield  {journal} {\bibinfo  {journal} {Sci.
  Rep.}\ }\textbf {\bibinfo {volume} {6}},\ \bibinfo {pages} {31624} (\bibinfo
  {year} {2016})}\BibitemShut {NoStop}%
\bibitem [{\citenamefont {Del~Sorbo}\ \emph {et~al.}(2017)\citenamefont
  {Del~Sorbo}, \citenamefont {Seipt}, \citenamefont {Blackburn}, \citenamefont
  {Thomas}, \citenamefont {Murphy}, \citenamefont {Kirk},\ and\ \citenamefont
  {Ridgers}}]{PhysRevA.96.043407}%
  \BibitemOpen
  \bibfield  {author} {\bibinfo {author} {\bibfnamefont {D.}~\bibnamefont
  {Del~Sorbo}}, \bibinfo {author} {\bibfnamefont {D.}~\bibnamefont {Seipt}},
  \bibinfo {author} {\bibfnamefont {T.~G.}\ \bibnamefont {Blackburn}}, \bibinfo
  {author} {\bibfnamefont {A.~G.~R.}\ \bibnamefont {Thomas}}, \bibinfo {author}
  {\bibfnamefont {C.~D.}\ \bibnamefont {Murphy}}, \bibinfo {author}
  {\bibfnamefont {J.~G.}\ \bibnamefont {Kirk}}, \ and\ \bibinfo {author}
  {\bibfnamefont {C.~P.}\ \bibnamefont {Ridgers}},\ }\href {\doibase
  10.1103/PhysRevA.96.043407} {\bibfield  {journal} {\bibinfo  {journal} {Phys.
  Rev. A}\ }\textbf {\bibinfo {volume} {96}},\ \bibinfo {pages} {043407}
  (\bibinfo {year} {2017})}\BibitemShut {NoStop}%
\bibitem [{\citenamefont {Sorbo}\ \emph {et~al.}(2018)\citenamefont {Sorbo},
  \citenamefont {Seipt}, \citenamefont {Thomas},\ and\ \citenamefont
  {Ridgers}}]{delSorbo2018polarization}%
  \BibitemOpen
  \bibfield  {author} {\bibinfo {author} {\bibfnamefont {D.~D.}\ \bibnamefont
  {Sorbo}}, \bibinfo {author} {\bibfnamefont {D.}~\bibnamefont {Seipt}},
  \bibinfo {author} {\bibfnamefont {A.~G.~R.}\ \bibnamefont {Thomas}}, \ and\
  \bibinfo {author} {\bibfnamefont {C.~P.}\ \bibnamefont {Ridgers}},\ }\href
  {http://stacks.iop.org/0741-3335/60/i=6/a=064003} {\bibfield  {journal}
  {\bibinfo  {journal} {Plasma Phys. Control. Fusion}\ }\textbf {\bibinfo
  {volume} {60}},\ \bibinfo {pages} {064003} (\bibinfo {year}
  {2018})}\BibitemShut {NoStop}%
\bibitem [{\citenamefont {Li}\ \emph {et~al.}(2019)\citenamefont {Li},
  \citenamefont {Shaisultanov}, \citenamefont {Hatsagortsyan}, \citenamefont
  {Wan}, \citenamefont {Keitel},\ and\ \citenamefont
  {Li}}]{PhysRevLett.122.154801}%
  \BibitemOpen
  \bibfield  {author} {\bibinfo {author} {\bibfnamefont {Y.-F.}\ \bibnamefont
  {Li}}, \bibinfo {author} {\bibfnamefont {R.}~\bibnamefont {Shaisultanov}},
  \bibinfo {author} {\bibfnamefont {K.~Z.}\ \bibnamefont {Hatsagortsyan}},
  \bibinfo {author} {\bibfnamefont {F.}~\bibnamefont {Wan}}, \bibinfo {author}
  {\bibfnamefont {C.~H.}\ \bibnamefont {Keitel}}, \ and\ \bibinfo {author}
  {\bibfnamefont {J.-X.}\ \bibnamefont {Li}},\ }\href {\doibase
  10.1103/PhysRevLett.122.154801} {\bibfield  {journal} {\bibinfo  {journal}
  {Phys. Rev. Lett.}\ }\textbf {\bibinfo {volume} {122}},\ \bibinfo {pages}
  {154801} (\bibinfo {year} {2019})}\BibitemShut {NoStop}%
\bibitem [{\citenamefont {Harding}\ and\ \citenamefont
  {Lai}(2006)}]{Harding2006}%
  \BibitemOpen
  \bibfield  {author} {\bibinfo {author} {\bibfnamefont {A.~K.}\ \bibnamefont
  {Harding}}\ and\ \bibinfo {author} {\bibfnamefont {D.}~\bibnamefont {Lai}},\
  }\href {\doibase 10.1088/0034-4885/69/9/R03} {\bibfield  {journal} {\bibinfo
  {journal} {Rep. Prog. Phys.}\ }\textbf {\bibinfo {volume} {69}},\ \bibinfo
  {pages} {2631} (\bibinfo {year} {2006})}\BibitemShut {NoStop}%
\bibitem [{\citenamefont {Uzdensky}\ and\ \citenamefont
  {Rightley}(2014)}]{uzdensky2014extreme}%
  \BibitemOpen
  \bibfield  {author} {\bibinfo {author} {\bibfnamefont {D.~A.}\ \bibnamefont
  {Uzdensky}}\ and\ \bibinfo {author} {\bibfnamefont {S.}~\bibnamefont
  {Rightley}},\ }\href {http://stacks.iop.org/0034-4885/77/i=3/a=036902}
  {\bibfield  {journal} {\bibinfo  {journal} {Rep. Prog. Phys.}\ }\textbf
  {\bibinfo {volume} {77}},\ \bibinfo {pages} {036902} (\bibinfo {year}
  {2014})}\BibitemShut {NoStop}%
\bibitem [{\citenamefont {Mahajan}\ \emph {et~al.}(2014)\citenamefont
  {Mahajan}, \citenamefont {Asenjo},\ and\ \citenamefont
  {Hazeltine}}]{Mahajan2014}%
  \BibitemOpen
  \bibfield  {author} {\bibinfo {author} {\bibfnamefont {S.~M.}\ \bibnamefont
  {Mahajan}}, \bibinfo {author} {\bibfnamefont {F.~A.}\ \bibnamefont {Asenjo}},
  \ and\ \bibinfo {author} {\bibfnamefont {R.~D.}\ \bibnamefont {Hazeltine}},\
  }\href {\doibase 10.1093/mnras/stu2381} {\bibfield  {journal} {\bibinfo
  {journal} {Mon. Not. R. Astron. Soc.}\ }\textbf {\bibinfo {volume} {446}},\
  \bibinfo {pages} {4112} (\bibinfo {year} {2014})}\BibitemShut {NoStop}%
\bibitem [{\citenamefont {Wolf}\ \emph {et~al.}(2001)\citenamefont {Wolf},
  \citenamefont {Awschalom}, \citenamefont {Buhrman}, \citenamefont {Daughton},
  \citenamefont {{Von Molnar}}, \citenamefont {Roukes}, \citenamefont
  {Chtchelkanova},\ and\ \citenamefont {Treger}}]{wolf2001spintronics}%
  \BibitemOpen
  \bibfield  {author} {\bibinfo {author} {\bibfnamefont {S.~A.}\ \bibnamefont
  {Wolf}}, \bibinfo {author} {\bibfnamefont {D.~D.}\ \bibnamefont {Awschalom}},
  \bibinfo {author} {\bibfnamefont {R.~A.}\ \bibnamefont {Buhrman}}, \bibinfo
  {author} {\bibfnamefont {J.~M.}\ \bibnamefont {Daughton}}, \bibinfo {author}
  {\bibfnamefont {S.}~\bibnamefont {{Von Molnar}}}, \bibinfo {author}
  {\bibfnamefont {M.~L.}\ \bibnamefont {Roukes}}, \bibinfo {author}
  {\bibfnamefont {A.~Y.}\ \bibnamefont {Chtchelkanova}}, \ and\ \bibinfo
  {author} {\bibfnamefont {D.~M.}\ \bibnamefont {Treger}},\ }\href@noop {}
  {\bibfield  {journal} {\bibinfo  {journal} {Science}\ }\textbf {\bibinfo
  {volume} {294}},\ \bibinfo {pages} {1488} (\bibinfo {year}
  {2001})}\BibitemShut {NoStop}%
\bibitem [{\citenamefont {Manfredi}\ and\ \citenamefont
  {Hervieux}(2007)}]{Manfredi-quantum-well}%
  \BibitemOpen
  \bibfield  {author} {\bibinfo {author} {\bibfnamefont {G.}~\bibnamefont
  {Manfredi}}\ and\ \bibinfo {author} {\bibfnamefont {P.-A.}\ \bibnamefont
  {Hervieux}},\ }\href@noop {} {\bibfield  {journal} {\bibinfo  {journal}
  {Appl. Phys. Lett.}\ }\textbf {\bibinfo {volume} {91}},\ \bibinfo {pages}
  {61108} (\bibinfo {year} {2007})}\BibitemShut {NoStop}%
\bibitem [{\citenamefont {Atwater}(2007)}]{atwater2007plasmonics}%
  \BibitemOpen
  \bibfield  {author} {\bibinfo {author} {\bibfnamefont {H.~A.}\ \bibnamefont
  {Atwater}},\ }\href@noop {} {\bibfield  {journal} {\bibinfo  {journal} {Sci.
  Am.}\ }\textbf {\bibinfo {volume} {296}},\ \bibinfo {pages} {56} (\bibinfo
  {year} {2007})}\BibitemShut {NoStop}%
\bibitem [{\citenamefont {Shukla}\ and\ \citenamefont
  {Eliasson}(2011)}]{Shukla-Eliasson-RMP}%
  \BibitemOpen
  \bibfield  {author} {\bibinfo {author} {\bibfnamefont {P.~K.}\ \bibnamefont
  {Shukla}}\ and\ \bibinfo {author} {\bibfnamefont {B.}~\bibnamefont
  {Eliasson}},\ }\href@noop {} {\bibfield  {journal} {\bibinfo  {journal} {Rev.
  Mod. Phys.}\ }\textbf {\bibinfo {volume} {83}},\ \bibinfo {pages} {885}
  (\bibinfo {year} {2011})}\BibitemShut {NoStop}%
\bibitem [{\citenamefont {Haas}(2011)}]{Haas-book}%
  \BibitemOpen
  \bibfield  {author} {\bibinfo {author} {\bibfnamefont {F.}~\bibnamefont
  {Haas}},\ }\href@noop {} {\emph {\bibinfo {title} {{Q}uantum {P}lasmas: An
  {H}ydrodynamic {A}pproach}}},\ Vol.~\bibinfo {volume} {65}\ (\bibinfo
  {publisher} {Springer},\ \bibinfo {address} {New York},\ \bibinfo {year}
  {2011})\BibitemShut {NoStop}%
\bibitem [{\citenamefont {Brodin}\ \emph {et~al.}(2008)\citenamefont {Brodin},
  \citenamefont {Marklund}, \citenamefont {Zamanian}, \citenamefont
  {Ericsson},\ and\ \citenamefont {Mana}}]{PhysRevLett.101.245002}%
  \BibitemOpen
  \bibfield  {author} {\bibinfo {author} {\bibfnamefont {G.}~\bibnamefont
  {Brodin}}, \bibinfo {author} {\bibfnamefont {M.}~\bibnamefont {Marklund}},
  \bibinfo {author} {\bibfnamefont {J.}~\bibnamefont {Zamanian}}, \bibinfo
  {author} {\bibfnamefont {{\AA}.}~\bibnamefont {Ericsson}}, \ and\ \bibinfo
  {author} {\bibfnamefont {P.~L.}\ \bibnamefont {Mana}},\ }\href {\doibase
  10.1103/PhysRevLett.101.245002} {\bibfield  {journal} {\bibinfo  {journal}
  {Phys. Rev. Lett.}\ }\textbf {\bibinfo {volume} {101}},\ \bibinfo {pages}
  {245002} (\bibinfo {year} {2008})}\BibitemShut {NoStop}%
\bibitem [{\citenamefont {Zamanian}\ \emph {et~al.}(2010)\citenamefont
  {Zamanian}, \citenamefont {Marklund},\ and\ \citenamefont
  {Brodin}}]{ZamanianNJP}%
  \BibitemOpen
  \bibfield  {author} {\bibinfo {author} {\bibfnamefont {J.}~\bibnamefont
  {Zamanian}}, \bibinfo {author} {\bibfnamefont {M.}~\bibnamefont {Marklund}},
  \ and\ \bibinfo {author} {\bibfnamefont {G.}~\bibnamefont {Brodin}},\
  }\href@noop {} {\bibfield  {journal} {\bibinfo  {journal} {New. J. Phys.}\
  }\textbf {\bibinfo {volume} {12}},\ \bibinfo {pages} {043019} (\bibinfo
  {year} {2010})}\BibitemShut {NoStop}%
\bibitem [{\citenamefont {Asenjo}\ \emph {et~al.}(2012)\citenamefont {Asenjo},
  \citenamefont {Zamanian}, \citenamefont {Marklund}, \citenamefont {Brodin},\
  and\ \citenamefont {Johansson}}]{asenjo2012semi}%
  \BibitemOpen
  \bibfield  {author} {\bibinfo {author} {\bibfnamefont {F.~A.}\ \bibnamefont
  {Asenjo}}, \bibinfo {author} {\bibfnamefont {J.}~\bibnamefont {Zamanian}},
  \bibinfo {author} {\bibfnamefont {M.}~\bibnamefont {Marklund}}, \bibinfo
  {author} {\bibfnamefont {G.}~\bibnamefont {Brodin}}, \ and\ \bibinfo {author}
  {\bibfnamefont {P.}~\bibnamefont {Johansson}},\ }\href@noop {} {\bibfield
  {journal} {\bibinfo  {journal} {New J. Phys.}\ }\textbf {\bibinfo {volume}
  {14}},\ \bibinfo {pages} {073042} (\bibinfo {year} {2012})}\BibitemShut
  {NoStop}%
\bibitem [{\citenamefont {Hurst}\ \emph {et~al.}(2014)\citenamefont {Hurst},
  \citenamefont {Morandi}, \citenamefont {Manfredi},\ and\ \citenamefont
  {Hervieux}}]{hurst2014semiclassical}%
  \BibitemOpen
  \bibfield  {author} {\bibinfo {author} {\bibfnamefont {J.}~\bibnamefont
  {Hurst}}, \bibinfo {author} {\bibfnamefont {O.}~\bibnamefont {Morandi}},
  \bibinfo {author} {\bibfnamefont {G.}~\bibnamefont {Manfredi}}, \ and\
  \bibinfo {author} {\bibfnamefont {P.-A.}\ \bibnamefont {Hervieux}},\
  }\href@noop {} {\bibfield  {journal} {\bibinfo  {journal} {Eur. Phys. J. D}\
  }\textbf {\bibinfo {volume} {68}},\ \bibinfo {pages} {176} (\bibinfo {year}
  {2014})}\BibitemShut {NoStop}%
\bibitem [{\citenamefont {Andreev}(2015)}]{andreev2015quantum}%
  \BibitemOpen
  \bibfield  {author} {\bibinfo {author} {\bibfnamefont {P.}~\bibnamefont
  {Andreev}},\ }\href@noop {} {\bibfield  {journal} {\bibinfo  {journal}
  {Physica A}\ }\textbf {\bibinfo {volume} {432}},\ \bibinfo {pages} {108}
  (\bibinfo {year} {2015})}\BibitemShut {NoStop}%
\bibitem [{\citenamefont {Andreev}(2017)}]{andreev2017kinetic}%
  \BibitemOpen
  \bibfield  {author} {\bibinfo {author} {\bibfnamefont {P.~A.}\ \bibnamefont
  {Andreev}},\ }\href@noop {} {\bibfield  {journal} {\bibinfo  {journal} {Phys.
  Plasmas}\ }\textbf {\bibinfo {volume} {24}},\ \bibinfo {pages} {022114}
  (\bibinfo {year} {2017})}\BibitemShut {NoStop}%
\bibitem [{\citenamefont {Ekman}\ \emph {et~al.}(2017)\citenamefont {Ekman},
  \citenamefont {Asenjo},\ and\ \citenamefont {Zamanian}}]{PhysRevE.96.023207}%
  \BibitemOpen
  \bibfield  {author} {\bibinfo {author} {\bibfnamefont {R.}~\bibnamefont
  {Ekman}}, \bibinfo {author} {\bibfnamefont {F.~A.}\ \bibnamefont {Asenjo}}, \
  and\ \bibinfo {author} {\bibfnamefont {J.}~\bibnamefont {Zamanian}},\ }\href
  {\doibase 10.1103/PhysRevE.96.023207} {\bibfield  {journal} {\bibinfo
  {journal} {Phys. Rev. E}\ }\textbf {\bibinfo {volume} {96}},\ \bibinfo
  {pages} {023207} (\bibinfo {year} {2017})}\BibitemShut {NoStop}%
\bibitem [{\citenamefont {Foldy}\ and\ \citenamefont
  {Wouthuysen}(1950)}]{foldy1950dirac}%
  \BibitemOpen
  \bibfield  {author} {\bibinfo {author} {\bibfnamefont {L.~L.}\ \bibnamefont
  {Foldy}}\ and\ \bibinfo {author} {\bibfnamefont {S.~A.}\ \bibnamefont
  {Wouthuysen}},\ }\href@noop {} {\bibfield  {journal} {\bibinfo  {journal}
  {Phys. Rev.}\ }\textbf {\bibinfo {volume} {78}},\ \bibinfo {pages} {29}
  (\bibinfo {year} {1950})}\BibitemShut {NoStop}%
\bibitem [{\citenamefont {Silenko}(2008)}]{Silenko2008}%
  \BibitemOpen
  \bibfield  {author} {\bibinfo {author} {\bibfnamefont {A.~J.}\ \bibnamefont
  {Silenko}},\ }\href {\doibase 10.1103/PhysRevA.77.012116} {\bibfield
  {journal} {\bibinfo  {journal} {Phys. Rev. A}\ }\textbf {\bibinfo {volume}
  {77}},\ \bibinfo {pages} {012116} (\bibinfo {year} {2008})}\BibitemShut
  {NoStop}%
\bibitem [{\citenamefont {Wigner}(1932)}]{wigner1932}%
  \BibitemOpen
  \bibfield  {author} {\bibinfo {author} {\bibfnamefont {E.~P.}\ \bibnamefont
  {Wigner}},\ }\href@noop {} {\bibfield  {journal} {\bibinfo  {journal} {Phys.
  Rev.}\ }\textbf {\bibinfo {volume} {40}},\ \bibinfo {pages} {749} (\bibinfo
  {year} {1932})}\BibitemShut {NoStop}%
\bibitem [{\citenamefont {Stratonovich}(1956)}]{stratonovich1956gauge}%
  \BibitemOpen
  \bibfield  {author} {\bibinfo {author} {\bibfnamefont {R.~L.}\ \bibnamefont
  {Stratonovich}},\ }\href@noop {} {\bibfield  {journal} {\bibinfo  {journal}
  {Sov. Phys. D}\ }\textbf {\bibinfo {volume} {1}},\ \bibinfo {pages} {414}
  (\bibinfo {year} {1956})}\BibitemShut {NoStop}%
\bibitem [{\citenamefont {Serimaa}\ \emph {et~al.}(1986)\citenamefont
  {Serimaa}, \citenamefont {Javanainen},\ and\ \citenamefont
  {Varr{\'o}}}]{serimaa}%
  \BibitemOpen
  \bibfield  {author} {\bibinfo {author} {\bibfnamefont {O.~T.}\ \bibnamefont
  {Serimaa}}, \bibinfo {author} {\bibfnamefont {J.}~\bibnamefont {Javanainen}},
  \ and\ \bibinfo {author} {\bibfnamefont {S.}~\bibnamefont {Varr{\'o}}},\
  }\href@noop {} {\bibfield  {journal} {\bibinfo  {journal} {Phys. Rev. A}\
  }\textbf {\bibinfo {volume} {33}},\ \bibinfo {pages} {2913} (\bibinfo {year}
  {1986})}\BibitemShut {NoStop}%
\bibitem [{\citenamefont {Case}(2008)}]{case2008wigner}%
  \BibitemOpen
  \bibfield  {author} {\bibinfo {author} {\bibfnamefont {W.~B.}\ \bibnamefont
  {Case}},\ }\href@noop {} {\bibfield  {journal} {\bibinfo  {journal} {Am. J.
  Phys.}\ }\textbf {\bibinfo {volume} {76}},\ \bibinfo {pages} {937} (\bibinfo
  {year} {2008})}\BibitemShut {NoStop}%
\bibitem [{\citenamefont {Shockley}\ and\ \citenamefont
  {James}(1967)}]{PhysRevLett.18.876}%
  \BibitemOpen
  \bibfield  {author} {\bibinfo {author} {\bibfnamefont {W.}~\bibnamefont
  {Shockley}}\ and\ \bibinfo {author} {\bibfnamefont {R.~P.}\ \bibnamefont
  {James}},\ }\href {\doibase 10.1103/PhysRevLett.18.876} {\bibfield  {journal}
  {\bibinfo  {journal} {Phys. Rev. Lett.}\ }\textbf {\bibinfo {volume} {18}},\
  \bibinfo {pages} {876} (\bibinfo {year} {1967})}\BibitemShut {NoStop}%
\bibitem [{\citenamefont {Shockley}(1968)}]{PhysRevLett.20.343}%
  \BibitemOpen
  \bibfield  {author} {\bibinfo {author} {\bibfnamefont {W.}~\bibnamefont
  {Shockley}},\ }\href {\doibase 10.1103/PhysRevLett.20.343} {\bibfield
  {journal} {\bibinfo  {journal} {Phys. Rev. Lett.}\ }\textbf {\bibinfo
  {volume} {20}},\ \bibinfo {pages} {343} (\bibinfo {year} {1968})}\BibitemShut
  {NoStop}%
\bibitem [{\citenamefont {Coleman}\ and\ \citenamefont
  {Van~Vleck}(1968)}]{PhysRev.171.1370}%
  \BibitemOpen
  \bibfield  {author} {\bibinfo {author} {\bibfnamefont {S.}~\bibnamefont
  {Coleman}}\ and\ \bibinfo {author} {\bibfnamefont {J.~H.}\ \bibnamefont
  {Van~Vleck}},\ }\href {\doibase 10.1103/PhysRev.171.1370} {\bibfield
  {journal} {\bibinfo  {journal} {Phys. Rev.}\ }\textbf {\bibinfo {volume}
  {171}},\ \bibinfo {pages} {1370} (\bibinfo {year} {1968})}\BibitemShut
  {NoStop}%
\bibitem [{\citenamefont {Babson}\ \emph {et~al.}(2009)\citenamefont {Babson},
  \citenamefont {Reynolds}, \citenamefont {Bjorkquist},\ and\ \citenamefont
  {Griffiths}}]{Babsonetal2009}%
  \BibitemOpen
  \bibfield  {author} {\bibinfo {author} {\bibfnamefont {D.}~\bibnamefont
  {Babson}}, \bibinfo {author} {\bibfnamefont {S.~P.}\ \bibnamefont
  {Reynolds}}, \bibinfo {author} {\bibfnamefont {R.}~\bibnamefont
  {Bjorkquist}}, \ and\ \bibinfo {author} {\bibfnamefont {D.~J.}\ \bibnamefont
  {Griffiths}},\ }\href {\doibase http://dx.doi.org/10.1119/1.3152712}
  {\bibfield  {journal} {\bibinfo  {journal} {Am. J. Phys.}\ }\textbf {\bibinfo
  {volume} {77}},\ \bibinfo {pages} {826} (\bibinfo {year} {2009})}\BibitemShut
  {NoStop}%
\bibitem [{\citenamefont {Lundin}\ and\ \citenamefont
  {Brodin}(2010)}]{lundin2010linearized}%
  \BibitemOpen
  \bibfield  {author} {\bibinfo {author} {\bibfnamefont {J.}~\bibnamefont
  {Lundin}}\ and\ \bibinfo {author} {\bibfnamefont {G.}~\bibnamefont
  {Brodin}},\ }\href@noop {} {\bibfield  {journal} {\bibinfo  {journal} {Phys.
  Rev. E}\ }\textbf {\bibinfo {volume} {82}},\ \bibinfo {pages} {056407}
  (\bibinfo {year} {2010})}\BibitemShut {NoStop}%
\bibitem [{\citenamefont {{Di Piazza}}\ \emph {et~al.}(2012)\citenamefont {{Di
  Piazza}}, \citenamefont {M{\"{u}}ller}, \citenamefont {Hatsagortsyan},\ and\
  \citenamefont {Keitel}}]{DiPiazza2012}%
  \BibitemOpen
  \bibfield  {author} {\bibinfo {author} {\bibfnamefont {A.}~\bibnamefont {{Di
  Piazza}}}, \bibinfo {author} {\bibfnamefont {C.}~\bibnamefont
  {M{\"{u}}ller}}, \bibinfo {author} {\bibfnamefont {K.~Z.}\ \bibnamefont
  {Hatsagortsyan}}, \ and\ \bibinfo {author} {\bibfnamefont {C.~H.}\
  \bibnamefont {Keitel}},\ }\href {\doibase 10.1103/RevModPhys.84.1177}
  {\bibfield  {journal} {\bibinfo  {journal} {Rev. Mod. Phys.}\ }\textbf
  {\bibinfo {volume} {84}},\ \bibinfo {pages} {1177} (\bibinfo {year}
  {2012})}\BibitemShut {NoStop}%
\bibitem [{\citenamefont {Burton}\ and\ \citenamefont
  {Noble}(2014)}]{Burton2014}%
  \BibitemOpen
  \bibfield  {author} {\bibinfo {author} {\bibfnamefont {D.~A.}\ \bibnamefont
  {Burton}}\ and\ \bibinfo {author} {\bibfnamefont {A.}~\bibnamefont {Noble}},\
  }\href {\doibase 10.1080/00107514.2014.886840} {\bibfield  {journal}
  {\bibinfo  {journal} {Contemp. Phys.}\ }\textbf {\bibinfo {volume} {55}},\
  \bibinfo {pages} {110} (\bibinfo {year} {2014})},\ \Eprint
  {http://arxiv.org/abs/1409.7707} {1409.7707} \BibitemShut {NoStop}%
\bibitem [{\citenamefont {Bhabha}\ and\ \citenamefont
  {Corben}(1941)}]{Bhabha1941}%
  \BibitemOpen
  \bibfield  {author} {\bibinfo {author} {\bibfnamefont {H.~J.}\ \bibnamefont
  {Bhabha}}\ and\ \bibinfo {author} {\bibfnamefont {H.~C.}\ \bibnamefont
  {Corben}},\ }\href {\doibase 10.1098/rspa.1941.0056} {\bibfield  {journal}
  {\bibinfo  {journal} {Proc. R. Soc. Lond. A}\ }\textbf {\bibinfo {volume}
  {178}},\ \bibinfo {pages} {273} (\bibinfo {year} {1941})}\BibitemShut
  {NoStop}%
\bibitem [{\citenamefont {Gonoskov}\ \emph {et~al.}(2015)\citenamefont
  {Gonoskov}, \citenamefont {Bastrakov}, \citenamefont {Efimenko},
  \citenamefont {Ilderton}, \citenamefont {Marklund}, \citenamefont {Meyerov},
  \citenamefont {Muraviev}, \citenamefont {Sergeev}, \citenamefont {Surmin},\
  and\ \citenamefont {Wallin}}]{gonoskov2015extended}%
  \BibitemOpen
  \bibfield  {author} {\bibinfo {author} {\bibfnamefont {A.}~\bibnamefont
  {Gonoskov}}, \bibinfo {author} {\bibfnamefont {S.}~\bibnamefont {Bastrakov}},
  \bibinfo {author} {\bibfnamefont {E.}~\bibnamefont {Efimenko}}, \bibinfo
  {author} {\bibfnamefont {A.}~\bibnamefont {Ilderton}}, \bibinfo {author}
  {\bibfnamefont {M.}~\bibnamefont {Marklund}}, \bibinfo {author}
  {\bibfnamefont {I.}~\bibnamefont {Meyerov}}, \bibinfo {author} {\bibfnamefont
  {A.}~\bibnamefont {Muraviev}}, \bibinfo {author} {\bibfnamefont
  {A.}~\bibnamefont {Sergeev}}, \bibinfo {author} {\bibfnamefont
  {I.}~\bibnamefont {Surmin}}, \ and\ \bibinfo {author} {\bibfnamefont
  {E.}~\bibnamefont {Wallin}},\ }\href@noop {} {\bibfield  {journal} {\bibinfo
  {journal} {Phys. Rev. E}\ }\textbf {\bibinfo {volume} {92}},\ \bibinfo
  {pages} {023305} (\bibinfo {year} {2015})}\BibitemShut {NoStop}%
\bibitem [{\citenamefont {Cole}\ \emph {et~al.}(2018)\citenamefont {Cole},
  \citenamefont {Behm}, \citenamefont {Gerstmayr}, \citenamefont {Blackburn},
  \citenamefont {Wood}, \citenamefont {Baird}, \citenamefont {Duff},
  \citenamefont {Harvey}, \citenamefont {Ilderton}, \citenamefont {Joglekar},
  \citenamefont {Krushelnick}, \citenamefont {Kuschel}, \citenamefont
  {Marklund}, \citenamefont {McKenna}, \citenamefont {Murphy}, \citenamefont
  {Poder}, \citenamefont {Ridgers}, \citenamefont {Samarin}, \citenamefont
  {Sarri}, \citenamefont {Symes}, \citenamefont {Thomas}, \citenamefont
  {Warwick}, \citenamefont {Zepf}, \citenamefont {Najmudin},\ and\
  \citenamefont {Mangles}}]{PhysRevX.8.011020}%
  \BibitemOpen
  \bibfield  {author} {\bibinfo {author} {\bibfnamefont {J.~M.}\ \bibnamefont
  {Cole}}, \bibinfo {author} {\bibfnamefont {K.~T.}\ \bibnamefont {Behm}},
  \bibinfo {author} {\bibfnamefont {E.}~\bibnamefont {Gerstmayr}}, \bibinfo
  {author} {\bibfnamefont {T.~G.}\ \bibnamefont {Blackburn}}, \bibinfo {author}
  {\bibfnamefont {J.~C.}\ \bibnamefont {Wood}}, \bibinfo {author}
  {\bibfnamefont {C.~D.}\ \bibnamefont {Baird}}, \bibinfo {author}
  {\bibfnamefont {M.~J.}\ \bibnamefont {Duff}}, \bibinfo {author}
  {\bibfnamefont {C.}~\bibnamefont {Harvey}}, \bibinfo {author} {\bibfnamefont
  {A.}~\bibnamefont {Ilderton}}, \bibinfo {author} {\bibfnamefont {A.~S.}\
  \bibnamefont {Joglekar}}, \bibinfo {author} {\bibfnamefont {K.}~\bibnamefont
  {Krushelnick}}, \bibinfo {author} {\bibfnamefont {S.}~\bibnamefont
  {Kuschel}}, \bibinfo {author} {\bibfnamefont {M.}~\bibnamefont {Marklund}},
  \bibinfo {author} {\bibfnamefont {P.}~\bibnamefont {McKenna}}, \bibinfo
  {author} {\bibfnamefont {C.~D.}\ \bibnamefont {Murphy}}, \bibinfo {author}
  {\bibfnamefont {K.}~\bibnamefont {Poder}}, \bibinfo {author} {\bibfnamefont
  {C.~P.}\ \bibnamefont {Ridgers}}, \bibinfo {author} {\bibfnamefont {G.~M.}\
  \bibnamefont {Samarin}}, \bibinfo {author} {\bibfnamefont {G.}~\bibnamefont
  {Sarri}}, \bibinfo {author} {\bibfnamefont {D.~R.}\ \bibnamefont {Symes}},
  \bibinfo {author} {\bibfnamefont {A.~G.~R.}\ \bibnamefont {Thomas}}, \bibinfo
  {author} {\bibfnamefont {J.}~\bibnamefont {Warwick}}, \bibinfo {author}
  {\bibfnamefont {M.}~\bibnamefont {Zepf}}, \bibinfo {author} {\bibfnamefont
  {Z.}~\bibnamefont {Najmudin}}, \ and\ \bibinfo {author} {\bibfnamefont
  {S.~P.~D.}\ \bibnamefont {Mangles}},\ }\href {\doibase
  10.1103/PhysRevX.8.011020} {\bibfield  {journal} {\bibinfo  {journal} {Phys.
  Rev. X}\ }\textbf {\bibinfo {volume} {8}},\ \bibinfo {pages} {011020}
  (\bibinfo {year} {2018})}\BibitemShut {NoStop}%
\bibitem [{\citenamefont {Poder}\ \emph {et~al.}(2018)\citenamefont {Poder},
  \citenamefont {Tamburini}, \citenamefont {Sarri}, \citenamefont {Di~Piazza},
  \citenamefont {Kuschel}, \citenamefont {Baird}, \citenamefont {Behm},
  \citenamefont {Bohlen}, \citenamefont {Cole}, \citenamefont {Corvan},
  \citenamefont {Duff}, \citenamefont {Gerstmayr}, \citenamefont {Keitel},
  \citenamefont {Krushelnick}, \citenamefont {Mangles}, \citenamefont
  {McKenna}, \citenamefont {Murphy}, \citenamefont {Najmudin}, \citenamefont
  {Ridgers}, \citenamefont {Samarin}, \citenamefont {Symes}, \citenamefont
  {Thomas}, \citenamefont {Warwick},\ and\ \citenamefont
  {Zepf}}]{PhysRevX.8.031004}%
  \BibitemOpen
  \bibfield  {author} {\bibinfo {author} {\bibfnamefont {K.}~\bibnamefont
  {Poder}}, \bibinfo {author} {\bibfnamefont {M.}~\bibnamefont {Tamburini}},
  \bibinfo {author} {\bibfnamefont {G.}~\bibnamefont {Sarri}}, \bibinfo
  {author} {\bibfnamefont {A.}~\bibnamefont {Di~Piazza}}, \bibinfo {author}
  {\bibfnamefont {S.}~\bibnamefont {Kuschel}}, \bibinfo {author} {\bibfnamefont
  {C.~D.}\ \bibnamefont {Baird}}, \bibinfo {author} {\bibfnamefont
  {K.}~\bibnamefont {Behm}}, \bibinfo {author} {\bibfnamefont {S.}~\bibnamefont
  {Bohlen}}, \bibinfo {author} {\bibfnamefont {J.~M.}\ \bibnamefont {Cole}},
  \bibinfo {author} {\bibfnamefont {D.~J.}\ \bibnamefont {Corvan}}, \bibinfo
  {author} {\bibfnamefont {M.}~\bibnamefont {Duff}}, \bibinfo {author}
  {\bibfnamefont {E.}~\bibnamefont {Gerstmayr}}, \bibinfo {author}
  {\bibfnamefont {C.~H.}\ \bibnamefont {Keitel}}, \bibinfo {author}
  {\bibfnamefont {K.}~\bibnamefont {Krushelnick}}, \bibinfo {author}
  {\bibfnamefont {S.~P.~D.}\ \bibnamefont {Mangles}}, \bibinfo {author}
  {\bibfnamefont {P.}~\bibnamefont {McKenna}}, \bibinfo {author} {\bibfnamefont
  {C.~D.}\ \bibnamefont {Murphy}}, \bibinfo {author} {\bibfnamefont
  {Z.}~\bibnamefont {Najmudin}}, \bibinfo {author} {\bibfnamefont {C.~P.}\
  \bibnamefont {Ridgers}}, \bibinfo {author} {\bibfnamefont {G.~M.}\
  \bibnamefont {Samarin}}, \bibinfo {author} {\bibfnamefont {D.~R.}\
  \bibnamefont {Symes}}, \bibinfo {author} {\bibfnamefont {A.~G.~R.}\
  \bibnamefont {Thomas}}, \bibinfo {author} {\bibfnamefont {J.}~\bibnamefont
  {Warwick}}, \ and\ \bibinfo {author} {\bibfnamefont {M.}~\bibnamefont
  {Zepf}},\ }\href {\doibase 10.1103/PhysRevX.8.031004} {\bibfield  {journal}
  {\bibinfo  {journal} {Phys. Rev. X}\ }\textbf {\bibinfo {volume} {8}},\
  \bibinfo {pages} {031004} (\bibinfo {year} {2018})}\BibitemShut {NoStop}%
\bibitem [{\citenamefont {Landau}\ and\ \citenamefont
  {Lifshitz}(1987)}]{landau1987classical}%
  \BibitemOpen
  \bibfield  {author} {\bibinfo {author} {\bibfnamefont {L.~D.}\ \bibnamefont
  {Landau}}\ and\ \bibinfo {author} {\bibfnamefont {E.~M.}\ \bibnamefont
  {Lifshitz}},\ }\href@noop {} {\emph {\bibinfo {title} {The {C}lassical
  {T}heory of {F}ields}}}\ (\bibinfo  {publisher} {Butterworth-Heinemann},\
  \bibinfo {address} {Oxford},\ \bibinfo {year} {1987})\BibitemShut {NoStop}%
\bibitem [{\citenamefont {Forger}\ and\ \citenamefont
  {R\"{o}mer}(2004)}]{forger2004currents}%
  \BibitemOpen
  \bibfield  {author} {\bibinfo {author} {\bibfnamefont {M.}~\bibnamefont
  {Forger}}\ and\ \bibinfo {author} {\bibfnamefont {H.}~\bibnamefont
  {R\"{o}mer}},\ }\href {\doibase https://doi.org/10.1016/j.aop.2003.08.011}
  {\bibfield  {journal} {\bibinfo  {journal} {Ann. Phys. (NY)}\ }\textbf
  {\bibinfo {volume} {309}},\ \bibinfo {pages} {306 } (\bibinfo {year}
  {2004})}\BibitemShut {NoStop}%
\bibitem [{\citenamefont {Jackson}(1999)}]{Jackson}%
  \BibitemOpen
  \bibfield  {author} {\bibinfo {author} {\bibfnamefont {J.~D.}\ \bibnamefont
  {Jackson}},\ }\href@noop {} {\emph {\bibinfo {title} {Classical
  {E}lectrodynamics}}},\ \bibinfo {edition} {3rd}\ ed.\ (\bibinfo  {publisher}
  {Wiley},\ \bibinfo {year} {1999})\BibitemShut {NoStop}%
\bibitem [{\citenamefont {Belinfante}(1940)}]{belinfante1940current}%
  \BibitemOpen
  \bibfield  {author} {\bibinfo {author} {\bibfnamefont {F.}~\bibnamefont
  {Belinfante}},\ }\href@noop {} {\bibfield  {journal} {\bibinfo  {journal}
  {Physica}\ }\textbf {\bibinfo {volume} {7}},\ \bibinfo {pages} {449}
  (\bibinfo {year} {1940})}\BibitemShut {NoStop}%
\bibitem [{\citenamefont {Rosenfeld}(1940)}]{rosenfeld1940tenseur}%
  \BibitemOpen
  \bibfield  {author} {\bibinfo {author} {\bibfnamefont {L.}~\bibnamefont
  {Rosenfeld}},\ }\href@noop {} {\bibfield  {journal} {\bibinfo  {journal}
  {Mém. Acad. Roy. Belg. Sci}\ }\textbf {\bibinfo {volume} {18}},\ \bibinfo
  {pages} {1} (\bibinfo {year} {1940})}\BibitemShut {NoStop}%
\bibitem [{\citenamefont {Goldman}(1977)}]{PhysRevD.15.1063}%
  \BibitemOpen
  \bibfield  {author} {\bibinfo {author} {\bibfnamefont {T.}~\bibnamefont
  {Goldman}},\ }\href {\doibase 10.1103/PhysRevD.15.1063} {\bibfield  {journal}
  {\bibinfo  {journal} {Phys. Rev. D}\ }\textbf {\bibinfo {volume} {15}},\
  \bibinfo {pages} {1063} (\bibinfo {year} {1977})}\BibitemShut {NoStop}%
\bibitem [{\citenamefont {Bargmann}\ \emph {et~al.}(1959)\citenamefont
  {Bargmann}, \citenamefont {Michel},\ and\ \citenamefont
  {Telegdi}}]{Bargmann1959}%
  \BibitemOpen
  \bibfield  {author} {\bibinfo {author} {\bibfnamefont {V.}~\bibnamefont
  {Bargmann}}, \bibinfo {author} {\bibfnamefont {L.}~\bibnamefont {Michel}}, \
  and\ \bibinfo {author} {\bibfnamefont {V.~L.}\ \bibnamefont {Telegdi}},\
  }\href {\doibase 10.1103/PhysRevLett.2.435} {\bibfield  {journal} {\bibinfo
  {journal} {Phys. Rev. Lett.}\ }\textbf {\bibinfo {volume} {2}},\ \bibinfo
  {pages} {435} (\bibinfo {year} {1959})}\BibitemShut {NoStop}%
\bibitem [{\citenamefont {{Page}}\ \emph {et~al.}(1996)\citenamefont {{Page}},
  \citenamefont {{Shibanov}},\ and\ \citenamefont
  {{Zavlin}}}]{page1996temperature}%
  \BibitemOpen
  \bibfield  {author} {\bibinfo {author} {\bibfnamefont {D.}~\bibnamefont
  {{Page}}}, \bibinfo {author} {\bibfnamefont {Y.~A.}\ \bibnamefont
  {{Shibanov}}}, \ and\ \bibinfo {author} {\bibfnamefont {V.~E.}\ \bibnamefont
  {{Zavlin}}},\ }in\ \href@noop {} {\emph {\bibinfo {booktitle}
  {Roentgenstrahlung from the Universe}}},\ \bibinfo {editor} {edited by\
  \bibinfo {editor} {\bibfnamefont {H.~U.}\ \bibnamefont {{Zimmermann}}},
  \bibinfo {editor} {\bibfnamefont {J.}~\bibnamefont {{Tr{\"u}mper}}}, \ and\
  \bibinfo {editor} {\bibfnamefont {H.}~\bibnamefont {{Yorke}}}}\ (\bibinfo
  {publisher} {Max Planck Institute for Extraterrestrial Physics},\ \bibinfo
  {address} {Garching},\ \bibinfo {year} {1996})\ pp.\ \bibinfo {pages}
  {173--174}\BibitemShut {NoStop}%
\bibitem [{\citenamefont {Mahajan}\ \emph {et~al.}(2013)\citenamefont
  {Mahajan}, \citenamefont {Machabeli}, \citenamefont {Osmanov},\ and\
  \citenamefont {Chkheidze}}]{mahajan2013ultra}%
  \BibitemOpen
  \bibfield  {author} {\bibinfo {author} {\bibfnamefont {S.}~\bibnamefont
  {Mahajan}}, \bibinfo {author} {\bibfnamefont {G.}~\bibnamefont {Machabeli}},
  \bibinfo {author} {\bibfnamefont {Z.}~\bibnamefont {Osmanov}}, \ and\
  \bibinfo {author} {\bibfnamefont {N.}~\bibnamefont {Chkheidze}},\ }\href@noop
  {} {\bibfield  {journal} {\bibinfo  {journal} {Sci. Rep.}\ }\textbf {\bibinfo
  {volume} {3}},\ \bibinfo {pages} {1262} (\bibinfo {year} {2013})}\BibitemShut
  {NoStop}%
\bibitem [{\citenamefont {Da~Costa}\ and\ \citenamefont
  {Kahn}(1991)}]{daCosta1991relativistic}%
  \BibitemOpen
  \bibfield  {author} {\bibinfo {author} {\bibfnamefont {A.}~\bibnamefont
  {Da~Costa}}\ and\ \bibinfo {author} {\bibfnamefont {F.}~\bibnamefont
  {Kahn}},\ }\href@noop {} {\bibfield  {journal} {\bibinfo  {journal} {Mon.
  Not. R. Astron. Soc.}\ }\textbf {\bibinfo {volume} {251}},\ \bibinfo {pages}
  {681} (\bibinfo {year} {1991})}\BibitemShut {NoStop}%
\bibitem [{\citenamefont {Lattimer}\ and\ \citenamefont
  {Prakash}(2004)}]{lattimer2004physics}%
  \BibitemOpen
  \bibfield  {author} {\bibinfo {author} {\bibfnamefont {J.~M.}\ \bibnamefont
  {Lattimer}}\ and\ \bibinfo {author} {\bibfnamefont {M.}~\bibnamefont
  {Prakash}},\ }\href@noop {} {\bibfield  {journal} {\bibinfo  {journal}
  {Science}\ }\textbf {\bibinfo {volume} {304}},\ \bibinfo {pages} {536}
  (\bibinfo {year} {2004})}\BibitemShut {NoStop}%
\bibitem [{\citenamefont {Shi}\ \emph {et~al.}(2018)\citenamefont {Shi},
  \citenamefont {Qin},\ and\ \citenamefont {Fisch}}]{shi2018laser}%
  \BibitemOpen
  \bibfield  {author} {\bibinfo {author} {\bibfnamefont {Y.}~\bibnamefont
  {Shi}}, \bibinfo {author} {\bibfnamefont {H.}~\bibnamefont {Qin}}, \ and\
  \bibinfo {author} {\bibfnamefont {N.~J.}\ \bibnamefont {Fisch}},\ }\href@noop
  {} {\bibfield  {journal} {\bibinfo  {journal} {Phys. Plasmas.}\ }\textbf
  {\bibinfo {volume} {25}},\ \bibinfo {pages} {055706} (\bibinfo {year}
  {2018})}\BibitemShut {NoStop}%
\bibitem [{\citenamefont {{Lindman}}(2010)}]{lindman2010laser}%
  \BibitemOpen
  \bibfield  {author} {\bibinfo {author} {\bibfnamefont {E.~L.}\ \bibnamefont
  {{Lindman}}},\ }\href {\doibase 10.1016/j.hedp.2010.02.002} {\bibfield
  {journal} {\bibinfo  {journal} {High Energy Density Phys.}\ }\textbf
  {\bibinfo {volume} {6}},\ \bibinfo {pages} {227} (\bibinfo {year}
  {2010})}\BibitemShut {NoStop}%
\bibitem [{\citenamefont {Wagner}\ \emph {et~al.}(2004)\citenamefont {Wagner},
  \citenamefont {Tatarakis}, \citenamefont {Gopal}, \citenamefont {Beg},
  \citenamefont {Clark}, \citenamefont {Dangor}, \citenamefont {Evans},
  \citenamefont {Haines}, \citenamefont {Mangles}, \citenamefont {Norreys},
  \citenamefont {Wei}, \citenamefont {Zepf},\ and\ \citenamefont
  {Krushelnick}}]{wagner2004laboratory}%
  \BibitemOpen
  \bibfield  {author} {\bibinfo {author} {\bibfnamefont {U.}~\bibnamefont
  {Wagner}}, \bibinfo {author} {\bibfnamefont {M.}~\bibnamefont {Tatarakis}},
  \bibinfo {author} {\bibfnamefont {A.}~\bibnamefont {Gopal}}, \bibinfo
  {author} {\bibfnamefont {F.~N.}\ \bibnamefont {Beg}}, \bibinfo {author}
  {\bibfnamefont {E.~L.}\ \bibnamefont {Clark}}, \bibinfo {author}
  {\bibfnamefont {A.~E.}\ \bibnamefont {Dangor}}, \bibinfo {author}
  {\bibfnamefont {R.~G.}\ \bibnamefont {Evans}}, \bibinfo {author}
  {\bibfnamefont {M.~G.}\ \bibnamefont {Haines}}, \bibinfo {author}
  {\bibfnamefont {S.~P.~D.}\ \bibnamefont {Mangles}}, \bibinfo {author}
  {\bibfnamefont {P.~A.}\ \bibnamefont {Norreys}}, \bibinfo {author}
  {\bibfnamefont {M.-S.}\ \bibnamefont {Wei}}, \bibinfo {author} {\bibfnamefont
  {M.}~\bibnamefont {Zepf}}, \ and\ \bibinfo {author} {\bibfnamefont
  {K.}~\bibnamefont {Krushelnick}},\ }\href@noop {} {\bibfield  {journal}
  {\bibinfo  {journal} {Phys. Rev. E}\ }\textbf {\bibinfo {volume} {70}},\
  \bibinfo {pages} {026401} (\bibinfo {year} {2004})}\BibitemShut {NoStop}%
\bibitem [{\citenamefont {Marklund}\ and\ \citenamefont
  {Shukla}(2006)}]{marklund2006nonlinear}%
  \BibitemOpen
  \bibfield  {author} {\bibinfo {author} {\bibfnamefont {M.}~\bibnamefont
  {Marklund}}\ and\ \bibinfo {author} {\bibfnamefont {P.~K.}\ \bibnamefont
  {Shukla}},\ }\href@noop {} {\bibfield  {journal} {\bibinfo  {journal} {Rev.
  Mod. Phys.}\ }\textbf {\bibinfo {volume} {78}},\ \bibinfo {pages} {591}
  (\bibinfo {year} {2006})}\BibitemShut {NoStop}%
\bibitem [{\citenamefont {Stefan}\ \emph {et~al.}(2011)\citenamefont {Stefan},
  \citenamefont {Zamanian}, \citenamefont {Brodin}, \citenamefont {Misra},\
  and\ \citenamefont {Marklund}}]{PhysRevE.83.036410}%
  \BibitemOpen
  \bibfield  {author} {\bibinfo {author} {\bibfnamefont {M.}~\bibnamefont
  {Stefan}}, \bibinfo {author} {\bibfnamefont {J.}~\bibnamefont {Zamanian}},
  \bibinfo {author} {\bibfnamefont {G.}~\bibnamefont {Brodin}}, \bibinfo
  {author} {\bibfnamefont {A.~P.}\ \bibnamefont {Misra}}, \ and\ \bibinfo
  {author} {\bibfnamefont {M.}~\bibnamefont {Marklund}},\ }\href {\doibase
  10.1103/PhysRevE.83.036410} {\bibfield  {journal} {\bibinfo  {journal} {Phys.
  Rev. E}\ }\textbf {\bibinfo {volume} {83}},\ \bibinfo {pages} {036410}
  (\bibinfo {year} {2011})}\BibitemShut {NoStop}%
\bibitem [{\citenamefont {Brodin}\ \emph {et~al.}(2010)\citenamefont {Brodin},
  \citenamefont {Misra},\ and\ \citenamefont
  {Marklund}}]{PhysRevLett.105.105004}%
  \BibitemOpen
  \bibfield  {author} {\bibinfo {author} {\bibfnamefont {G.}~\bibnamefont
  {Brodin}}, \bibinfo {author} {\bibfnamefont {A.~P.}\ \bibnamefont {Misra}}, \
  and\ \bibinfo {author} {\bibfnamefont {M.}~\bibnamefont {Marklund}},\ }\href
  {\doibase 10.1103/PhysRevLett.105.105004} {\bibfield  {journal} {\bibinfo
  {journal} {Phys. Rev. Lett.}\ }\textbf {\bibinfo {volume} {105}},\ \bibinfo
  {pages} {105004} (\bibinfo {year} {2010})}\BibitemShut {NoStop}%
\bibitem [{\citenamefont {Nikishov}\ and\ \citenamefont
  {Ritus}(1964{\natexlab{a}})}]{nikishov1964quantumI}%
  \BibitemOpen
  \bibfield  {author} {\bibinfo {author} {\bibfnamefont {A.}~\bibnamefont
  {Nikishov}}\ and\ \bibinfo {author} {\bibfnamefont {V.}~\bibnamefont
  {Ritus}},\ }\href@noop {} {\bibfield  {journal} {\bibinfo  {journal} {Sov.
  Phys. JETP}\ }\textbf {\bibinfo {volume} {19}},\ \bibinfo {pages} {529}
  (\bibinfo {year} {1964}{\natexlab{a}})}\BibitemShut {NoStop}%
\bibitem [{\citenamefont {Nikishov}\ and\ \citenamefont
  {Ritus}(1964{\natexlab{b}})}]{nikishov1964quantumII}%
  \BibitemOpen
  \bibfield  {author} {\bibinfo {author} {\bibfnamefont {A.}~\bibnamefont
  {Nikishov}}\ and\ \bibinfo {author} {\bibfnamefont {V.}~\bibnamefont
  {Ritus}},\ }\href@noop {} {\bibfield  {journal} {\bibinfo  {journal} {Sov.
  Phys. JETP}\ }\textbf {\bibinfo {volume} {19}},\ \bibinfo {pages} {1191}
  (\bibinfo {year} {1964}{\natexlab{b}})}\BibitemShut {NoStop}%
\bibitem [{\citenamefont {Harvey}\ and\ \citenamefont
  {Marklund}(2012)}]{harvey2012radiation}%
  \BibitemOpen
  \bibfield  {author} {\bibinfo {author} {\bibfnamefont {C.}~\bibnamefont
  {Harvey}}\ and\ \bibinfo {author} {\bibfnamefont {M.}~\bibnamefont
  {Marklund}},\ }\href@noop {} {\bibfield  {journal} {\bibinfo  {journal}
  {Phys. Rev. A}\ }\textbf {\bibinfo {volume} {85}},\ \bibinfo {pages} {013412}
  (\bibinfo {year} {2012})}\BibitemShut {NoStop}%
\bibitem [{\citenamefont {Stefan}\ and\ \citenamefont
  {Brodin}(2013)}]{stefan2013linear}%
  \BibitemOpen
  \bibfield  {author} {\bibinfo {author} {\bibfnamefont {M.}~\bibnamefont
  {Stefan}}\ and\ \bibinfo {author} {\bibfnamefont {G.}~\bibnamefont
  {Brodin}},\ }\href@noop {} {\bibfield  {journal} {\bibinfo  {journal} {Phys.
  Plasmas}\ }\textbf {\bibinfo {volume} {20}},\ \bibinfo {pages} {012114}
  (\bibinfo {year} {2013})}\BibitemShut {NoStop}%
\bibitem [{\citenamefont {Dittrich}\ and\ \citenamefont
  {Gies}(2000)}]{dittrich2000probing}%
  \BibitemOpen
  \bibfield  {author} {\bibinfo {author} {\bibfnamefont {W.}~\bibnamefont
  {Dittrich}}\ and\ \bibinfo {author} {\bibfnamefont {H.}~\bibnamefont
  {Gies}},\ }\href@noop {} {\emph {\bibinfo {title} {Probing the {Q}uantum
  {V}acuum: {P}erturbative {E}ffective {A}ction {A}pproach in {Q}uantum
  {E}lectrodynamics and {I}ts {A}pplication}}},\ Vol.\ \bibinfo {volume} {166}\
  (\bibinfo  {publisher} {Springer},\ \bibinfo {year} {2000})\BibitemShut
  {NoStop}%
\bibitem [{Note1()}]{Note1}%
  \BibitemOpen
  \bibinfo {note} {Particle contributions enter a dispersion relation
  proportional to a factor $\omega _p^2/ \omega ^2$, and hence are suppressed
  for low densities or high frequencies.}\BibitemShut {Stop}%
\bibitem [{\citenamefont {Haas}\ \emph {et~al.}(2010)\citenamefont {Haas},
  \citenamefont {Zamanian}, \citenamefont {Marklund},\ and\ \citenamefont
  {Brodin}}]{haas2010fluid}%
  \BibitemOpen
  \bibfield  {author} {\bibinfo {author} {\bibfnamefont {F.}~\bibnamefont
  {Haas}}, \bibinfo {author} {\bibfnamefont {J.}~\bibnamefont {Zamanian}},
  \bibinfo {author} {\bibfnamefont {M.}~\bibnamefont {Marklund}}, \ and\
  \bibinfo {author} {\bibfnamefont {G.}~\bibnamefont {Brodin}},\ }\href@noop {}
  {\bibfield  {journal} {\bibinfo  {journal} {New J. Phys.}\ }\textbf {\bibinfo
  {volume} {12}},\ \bibinfo {pages} {073027} (\bibinfo {year}
  {2010})}\BibitemShut {NoStop}%
\bibitem [{\citenamefont {Stefan}(2014)}]{stefan2014models}%
  \BibitemOpen
  \bibfield  {author} {\bibinfo {author} {\bibfnamefont {M.}~\bibnamefont
  {Stefan}},\ }\emph {\bibinfo {title} {On Models of Quantum Plasmas and their
  Nonlinear Implications}},\ \href@noop {} {Ph.D. thesis},\ \bibinfo  {school}
  {Ume{\aa} {U}niversity} (\bibinfo {year} {2014})\BibitemShut {NoStop}%
\bibitem [{Note2()}]{Note2}%
  \BibitemOpen
  \bibinfo {note} {Discussed in Paper I~\cite {PhysRevE.96.023207}, see also
  references therein.}\BibitemShut {Stop}%
\end{thebibliography}%
	
\end{document}